\documentclass[twocolumn]{aastex62}
\newcommand{\ha}{H$\alpha$}    
\newcommand{\hb}{H$\beta$}    
\newcommand{\sii}{[S~{\sc ii}]}
\newcommand{\oiii}{[O~{\sc iii}]}

\newcommand{\nii}{[N~{\sc ii}]}

\newcommand{\newcode}{{$\rm M^3$}}
\newcommand{\mappings}{{\sc mappings~v}}

\newcommand{\hiireg}{{H{\sc ii}}}

\graphicspath{{./}{figures/}}
\usepackage{CJK}
\usepackage{amsmath}

\received{January 1, 2018}
\revised{January 7, 2018}
\accepted{\today}
\submitjournal{ApJ}

\shorttitle{Messenger Monte-Carlo MAPPINGS}
\shortauthors{Jin et al.}

\begin{document}
\begin{CJK*}{UTF8}{gbsn}

\title{Messenger Monte-Carlo MAPPINGS V (\newcode ) -- A self-consistent three-dimensional photoionization code}

\correspondingauthor{Yifei Jin}
\email{Yifei.Jin@anu.edu.au}

\author[0000-0003-0401-3688]{Yifei Jin (金刈非)}
\affil{Research School for Astronomy \& Astrophysics, Australian National University, Canberra, Australia, 2611}
\affiliation{ARC Centre of Excellence for All Sky Astrophysics in 3 Dimensions (ASTRO 3D)}

\author[0000-0001-8152-3943]{Lisa J. Kewley}
\affiliation{Research School for Astronomy \& Astrophysics, Australian National University, Canberra, Australia, 2611}
\affiliation{ARC Centre of Excellence for All Sky Astrophysics in 3 Dimensions (ASTRO 3D)}

\author[0000-0002-6620-7421]{Ralph Sutherland}
\affiliation{Research School for Astronomy \& Astrophysics, Australian National University, Canberra, Australia, 2611}

\begin{abstract}

The Messenger Interface Monte-Carlo Mappings V (\newcode) is a photoionization code adopting the fully self-consistent Monte-Carlo radiative transfer technique, which presents a major advance over previous photoionization models with simple geometries.
\newcode\ is designed for modeling nebulae in arbitrary three-dimensional geometries.
In this paper, we describe the Monte-Carlo radiative transfer technique and the microphysics implemented in \newcode , including the photoionization, collisional ionization, the free-free and free-bound recombination, and two-photon radiation.
We put \newcode\ through the Lexington/Meudon benchmarks to test the reliability of the new code.
We apply \newcode\ to three \hiireg\ region models with fiducial geometries, demonstrating that \newcode\ is capable of dealing with nebulae with complex geometries.
\newcode\ is a promising tool for understanding emission-line behavior in the era of SDSS-V/LVM and JWST, which will provide high-quality data of spatially-resolved nearby \hiireg\ regions and highly turbulent local and high-redshift \hiireg\ regions.

\end{abstract}


\section{Introduction}\label{sec:intro}

Modeling nebular emission-line regions is vital for the interpretation of spectroscopic data.
Comparing the modeled emission-line spectra with observations can determine the central power source in galaxies, such as star formation, AGN or shocks \citep{Baldwin-1981,Veilleux-1987,Osterbrock-1992,Kewley-2001,D'Agostino-2019,Byler-2020}. 
Accurate models of emission-line ratios provide the fundamental properties of the interstellar medium (ISM), including the metallicity \citep{Tremonti-2004,Yuan-2013,Yabe-2015,Sanders-2020}, the pressure and the electron density of the ionized gas \citep{Kaasinen-2017,Kewley-2019,Harshan-2020,Davies-2021}. 

Photoionization codes are fundamental tools to model emission-line regions \citep{Ferland-1998,Netzer-1993,Sutherland-1993,Ercolano-2003,Morisset-2006}. 
These models combine atomic data, radiative transfer processes and the physics of the ISM.
\cite{Ferland-1995} summarize the features of photoionization models.
The architectures are similar in most photoionization codes. 
Nebular models treat the ISM as a plane parallel slab or sphere, dividing the ISM into a series of zones which satisfy ionization and thermal equilibrium \citep{Dopita-2000,Groves-2004}.
The solution of the radiative transfer equation in each zone is based on the local ionization conditions.
The ionization states in each zone are determined by balancing ionization and recombination, including the processes of photoionization, collisional ionization, radiative and dielectronic recombinations. 
The thermal structure of the nebula is determined by the balance between cooling and heating processes \citep{Sutherland-1993}.
The coefficients for ionization and recombination rates are generated from the fundamental atomic data \citep[See][for a review]{Stasinska-2002}.

The treatment of the diffuse radiation is sophisticated in photoionization models.
Approximations are always adopted in photoionization codes because of the complexity of calculating the diffuse radiation in radiative transfer processes \citep{Wood-2004}.
The outward-only approximation is predominantly implemented in photoionization codes \citep{Ferland-1998,Netzer-1993,Binette-1985}, which assumes that the locally produced diffuse radiations follow the outward direction of incident photons. 
The limitation of the outward-only approximation is that too much energy is trapped in the inner part of photoionized regions, creating an unrealistic temperature structure \citep{Blandford-1990}.
Alternatively, \cite{Harrington-1968} and \cite{Rubin-1991} adopted a full treatment of diffuse photons where the diffuse radiations are split into an outward stream and a backward stream, whose contributions are iteratively calculated across the photoionized region.
Currently, the full treatment of diffuse radiation only suits simple nebular geometries, like a spherical or axisymmetric geometry.

Simplification of the nebular geometry is another problem in most photoionization codes.
The geometry is usually assumed to be spherical or plane-parallel for the analysis of the diagnostic emission-lines \citep{Kobulnicky-2004,Levesque-2010} and of the power sources in galaxies \citep{Kewley-2001}.
However, the realistic structure of \hiireg\ regions is too complex to be simply described by spherical or plane-parallel geometry.
Detailed observations of the Orion Nebula show a concave structure with a bright dense ionized bar in the foreground of a veil of low-dense ionized gas \citep{Zuckerman-1973,O'Dell-2009}.
In other nearby nebulae, like the ``Pillars of Creation'', a large amount of filamentary structures exist \citep{Schneider-2016} due to the joint effects from ISM turbulence and stellar radiation \citep{Gritschneder-2010,Tremblin-2013}.
These nebular structures complicate the density and temperature structures of \hiireg\ regions, altering the fluxes of diagnostic emission-lines, like \sii$\lambda\lambda$6717,31, \oiii$\lambda$5007, \nii$\lambda$6583 and \ha\ \citep{Kewley-2019}.

Monte Carlo radiative transfer (MCRT) provides much promise in modeling realistic photoionized regions in three dimensions. 
One pronounced feature of MCRT is that it can simulate the genuine radiative transfer process in ISM with an arbitrary geometry.
The diffuse radiation in MCRT are fully treated such that the diffuse photons are produced based on the local ISM ionization conditions and are emitted isotropically.
The MCRT technique has been successfully applied in some three-dimensional photoionization simulations \citep{Ercolano-2003}. 

\newcode\ is a new self-consistent 3D photoionization code, combining the Monte Carlo radiative transfer technique with the well-known \mappings\ photoionization code.
The purpose of \newcode\ is to produce reliable diagnostic emission-lines in the nebulae with arbitrary geometry. 
\mappings\ \citep{Binette-1985,Sutherland-1993,Sutherland-2018} is a large photoionization code including 30 elements and 80,000 cooling and recombination lines generated from CHIANTI v.8.0 atomic database \citep{Del Zanna-2015}.
The code self-consistently calculates nebular cooling and heating processes as well as the complex physics of dust grains.

Compared to other three-dimensional photoionization codes \citep{Ercolano-2003,Wood-2004,Vandenbroucke-2018}, \newcode\ has a more comprehensive consideration of microphysics in the ISM, including cooling and heating. 
\newcode\ inherits engines from \mappings\ to solve ionization and thermal balance, which are well tested in both observational and theoretical models.

This paper is structured as follows. 
In Section~\ref{sec:mcrt}, we briefly describe the Monte Carlo radiative transfer technique used in our \newcode\ code.
Sections~\ref{sec:ionrec} and~\ref{sec:heatcool} describe the microphysics we adopt.
In Section~\ref{sec:benchmark}, we demonstrate the reliability of \newcode\ by comparing the outputs of benchmark cases between \newcode\ and other photoionization codes.
We present two nebular models with complex geometries produced by \newcode\ in Section~\ref{sec:models}.
In Section~\ref{sec:discussion}, we discuss the capabilities of \newcode .
A summary is given in Section~\ref{sec:summary}.

\section{The architecture of \newcode }

\begin{figure*}
  \centering
  \includegraphics[width=1\textwidth]{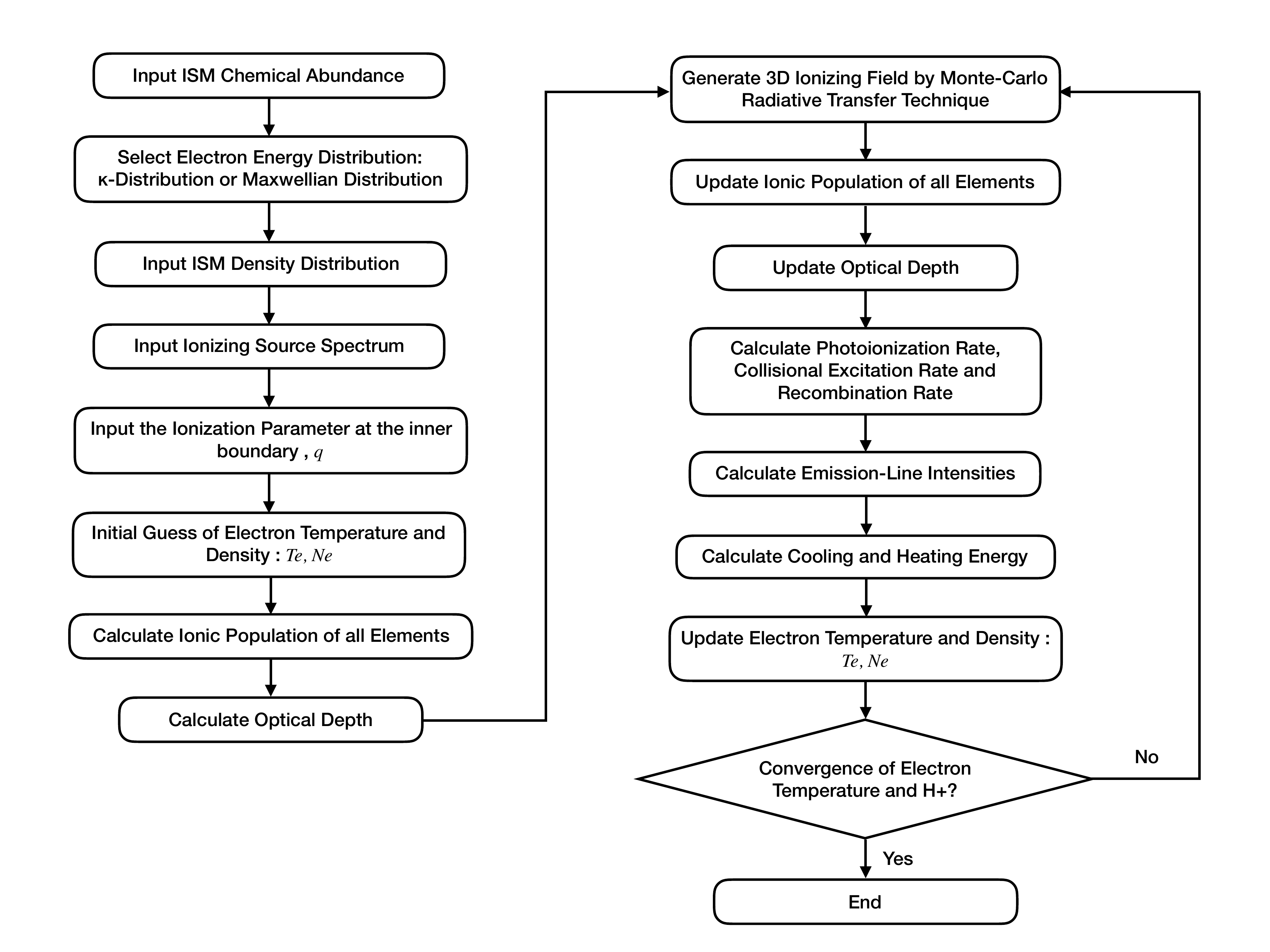}
  \caption{\newcode\ working flowchart. Each step indicates the operational function during the photoionization modeling.}\label{fig1}
\end{figure*}

The \newcode\ code is built on two major steps: the Monte Carlo radiative transfer (MCRT), and the thermal and ionization balance calculation.
The Monte Carlo radiative transfer is implemented to construct a three-dimensional ionization field of the photoionized region, based on which \newcode\ solves the local balance between ionization and recombination, cooling and heating processes.
A set of electron temperature, electron density and ionic fraction of chemical elements are updated, feeding into a new round of Monte Carlo radiative transfer, until the convergence of the electron temperature and the ionic fractions of hydrogen. 
The detailed working flowchart is presented in Figure~\ref{fig1}.
 
\section{Monte Carlo radiative transfer}\label{sec:mcrt}

The main principle of Monte Carlo treatments of radiative transfer is to take photons as the calculation quanta to simulate the local physical processes of ionization and recombination. 
In the \newcode\ code, we implement the Monte Carlo radiative transfer method built by \cite{Lucy-1999}, in which photons of the same frequency, $\nu$, are grouped into energy packets, $\varepsilon (\nu)$, so that
\begin{equation}\label{eq1}
 \varepsilon (\nu)=nh\nu,
\end{equation}
where $n$ is the number of photons in each energy packet.
The use of energy packets rather than individual photons as the calculation quanta improves the computational efficiency \citep{Och-1998,Lucy-1999}. 
To increase the accuracy of the ionizing spectral sampling, we adopt a frequency-dependent energy packet given by
\begin{equation}\label{eq1.1} 
 \varepsilon_{\nu}=L_{\nu}\Delta t/N , 
\end{equation}
where $L_{\nu}$ is the luminosity of the source at frequency $\nu$, $\Delta t$ is the time of the simulation, and $N$ is the number of energy packets used at each frequency. 
The direction vectors, ($n_x$,$n_y$,$n_z$), of energy packets are created isotropically in Cartesian coordinates by
\begin{equation}\label{eq2}
\begin{split}
n_z=2\alpha-1\\
\theta=\pi(2\beta-1)\\
n_x=\sqrt{(1-n^2_z)}cos\theta\\
n_y=\sqrt{(1-n^2_z)}sin\theta
\end{split}
\end{equation}
where $\alpha$ and $\beta$ are random numbers generated between [0,1]. 

Each energy packet is then launched from the emitting source, travels through the entire simulated domain and contributes to the local radiation field of the cells it travels through.
The mean intensity of the radiation field in each cell is the sum of all passing energy packets, which is given by 
\begin{equation}\label{eq6}
 J(\nu)=\frac{1}{4\pi}\sum_{N^{\prime}}\left(\frac{\varepsilon_{\nu}}{\Delta t}\frac{l}{V}\right),
\end{equation}
where $J(\nu)$ is the mean intensity of radiation field, $N^{\prime}$ is the number of energy packets passing through the cell, $V$ and $l$ are the volume of the cell and the displacement of each packet. 
The estimated mean intensity is then fed to the local ionization and recombination processes for calculating the temperature and ionization status.

\subsection{The trajectories of energy packets}

The trajectory of each energy packet must be tracked in order to determine the locations of the absorption events followed by re-emissions of each packet during its journey through the nebula.
In a ``density bounded" nebula, the trajectories of packets terminate at the boundary of the nebula. 
In a ``radiation bounded" nebula, the packets end their journey after undergoing a specific number of the ``absorption -- re-emission'' loops in the nebula.

We adopt the ``cell-by-cell'' tracking strategy suggested by \cite{Lucy-1999}, which follows each energy packet cell by cell along its trajectory in concert to check the occurrence of absorption events. 
There is an alternative method suggested by \cite{Harries-1997} that is to first calculate the probability of the occurrence of absorption events as a function of the distance to the central ionizing source. The location of the absorption event is then searched for along the radius of the nebula by comparing a random number against the pre-calculated probability function.
However, this method is less efficient than \cite{Lucy-1999} method because it requires searching routines in the computations \citep{Ercolano-2003}.

\subsection{Absorption and re-emission}

The absorption is determined by comparing the random optical depth, $\tau_{p}(\nu)$, with the analytical optical depth, $\tau_{l}(\nu)$, derived from the physical displacement $l$.
The $\tau_{l}$ is given by
\begin{equation}\label{eq3}
\tau_{l}(\nu)=\kappa(\nu)\rho l,
\end{equation}
where $\kappa(\nu)$ and $\rho$ are the absorption coefficient and the volume density of the local cell. 
If $\tau_{l}$ is smaller than the random optical depth $\tau_{p}$, then the packet moves along its initial direction to its next stopping-point in the adjacent cell, where a new $\tau_{p}$ is assigned. 
The stopping-point is the location in each cell where the optical depth is updated. 
The distance between two stopping points is $l$.
If $\tau_{l}$ is larger than $\tau_{p}$, the packet moves the distance of $\tau_{p}/(\kappa\rho)$, where the absorption event and the follow-up re-emission occur.
In this scheme, the $\tau_{p}(\nu)$ is regulated by a logarithmic formula given by,
\begin{equation}\label{eq4}
\tau_{p}(\nu)=-{\rm ln}(p),
\end{equation}
where $p$ is a random number between [0,1].
The regulation of $\tau_p$ guarantees the statistical fluxes of the ionizing spectrum follow an exponential decline along the distance from the central source.

The energy packet is re-emitted in situ immediately once absorption occurs. 
The frequency of the re-emitted energy packet is determined by the probability density function (PDF) derived from the local emissivity distribution.
The probability density function is given by,
\begin{equation}\label{eq5}
 p (\nu)=\frac{I_\nu}{\int_{\nu_{min}}^{\nu_{max}}{I_{\nu^{\prime}}} d\nu^{\prime}},
\end{equation}
where $I_\nu$ is the emissivity distribution of the diffuse radiation, $\nu_{min}$ and $\nu_{max}$ bracket the frequency range of the diffuse ionizing spectrum, and $p (\nu)$ denotes the probability of an energy packet created at the frequency $\nu$. 
The local emissivity includes the recombination emission-lines and the continuum recombination.

\section{Ionization and recombination}\label{sec:ionrec}

\subsection{Photoionization and Collisional Ionization}

In \newcode , the photoionization cross section is calculated using the method suggested by \cite{Seaton-1958}, with the photoionization data given by \cite{Raymond-1979}, \cite{Osterbrock-1989} and \cite{Gould-1991}.
We also consider the Auger transition by including the modification factor from \cite{Weisheit-1974}.

The collisional ionization rates are calculated from the electron energy distribution.
The Maxwellian distribution is predominantly adopted in \hiireg\ region models where the gas is assumed to be completely thermalized \citep{Spitzer-1941, Draine-2018}.
Recently, the $\kappa$ electron energy distribution has been found in interplanetary plasmas \citep{Pierrard-2010} and used to account for the observed line ratios in \hiireg\ regions \citep{Binette-2012,Nicholls-2012,Nicholls-2013}.

The electron energy distribution can be chosen to be a Maxwellian distribution or a $\kappa$-distribution in \newcode .
Under the assumption of a Maxwellian distribution, the collisional ionization rates are calculated based on the method proposed by \cite{Arnaud-1985} and \cite{Younger-1981}.  
This method uses a five parameter fit to the collisional cross section and derives an expression for the integral over a Maxwellian velocity distribution in electron energy.
For the $\kappa$-distribution, we add a theoretical correction factor to the photoionization rate for each atomic shell \citep{Nicholls-2012,Nicholls-2013}.

\subsection{Recombination}

We include radiative recombination and dielectronic recombination for both hydrogenic and non-hydrogenic ions.
The recombination rates for hydrogenic ions and non-hydrogenic ions are calculated separately.
We use the power-law fits proposed by \cite{Aldrovandi-1973} to calculate the recombination rates for all non-hydrogenic ions.

The recombination rates for hydrogenic ions are treated carefully because hydrogen is the dominant species in the ISM and has a critical impact on nebular properties.
We follow the strategy proposed by \cite{Sutherland-1993}, which provides a temperature-dependent calculation of recombination rates.
The \cite{Seaton-1959} recombination rates are used when $log(T/Z^{2}) < 6.0$, where $T$ is the temperature and $Z$ is the nuclear charge number.
When $log(T/Z^{2})\geq 6.0$, a ``quantum correction'' factor \citep{Gaunt-1930} is added to the recombination rates in order to avoid the divergence of the \cite{Seaton-1959} calculation in this temperature range.

\subsection{Continuum Radiation and Line Radiation}

The continuum radiation in \newcode\ consists of three parts: free-free, free-bound and two-photon radiation. 
We adopt the Gaunt factors given by \cite{Gronenschild-1978} for free-free and two-photon radiation.
The Gaunt factors for free-bound radiation are determined by the expression in \cite{Mewe-1986}.

The line radiation in \newcode\ includes resonance lines, forbidden lines, inter-system lines and fine-structure lines.
The hydrogen and helium resonance lines are given as the linear combination of the Case A and Case B emissivities.
The remaining resonance lines are calculated based on the expressions given by \cite{Mewe-1981}.
In the current version of \newcode, we assume that the resonance lines have the same cross section as the continuum emission.
The forbidden-line emissivities are computed by treating the atoms and ions as a five-level system. 
Because the forbidden lines are always optically thin, we assume that the forbidden-line emissions carry away energies from the ISM without further interaction.
Finally, the inter-system lines and fine-structure lines are treated as the radiation from a two-level system using the method suggested in \cite{Binette-1982}.

\section{Heating and Cooling Processes}\label{sec:heatcool}

\newcode\ inherits the cooling and heating calculations from \mappings, which are described in detail in \cite{Sutherland-1993}.
Briefly, the net cooling function is defined as the difference between the energy lost through cooling processes and the energy gained through heating processes.
The collisional line, free-free and two-photon radiations are the major cooling sources. 
The heating mechanisms consist of photoionization heating, collisional ionization and Compton heating. 
The radiation from recombination processes may contribute to either heating or cooling effects, depending on the electron energy distribution.

\section{Decomposition of Simulated Domains}\label{sec:mpi}

High spatial and frequency resolved Monte-Carlo photoionization models require CPUs with large memory.
In order to run photoionization models with large grids, we divide the simulated domain into a series of equal-sized blocks, which are distributed to separate CPU-cores on supercomputers.
These blocks contain the same number of cells and the adjacent blocks share the same cell wall.
Each block contains the data to compute the local thermal and ionization status independently.
The frequency and direction vector of energy packets, are transferred between adjacent blocks through the MPI (Message Passing Interface) message passing standard.

\section{Testing $\rm M^{3}$ with the $Lexington/Meudon$ Standard Models}\label{sec:benchmark}


\begin{figure*}
  \centering
  \includegraphics[width=1\textwidth]{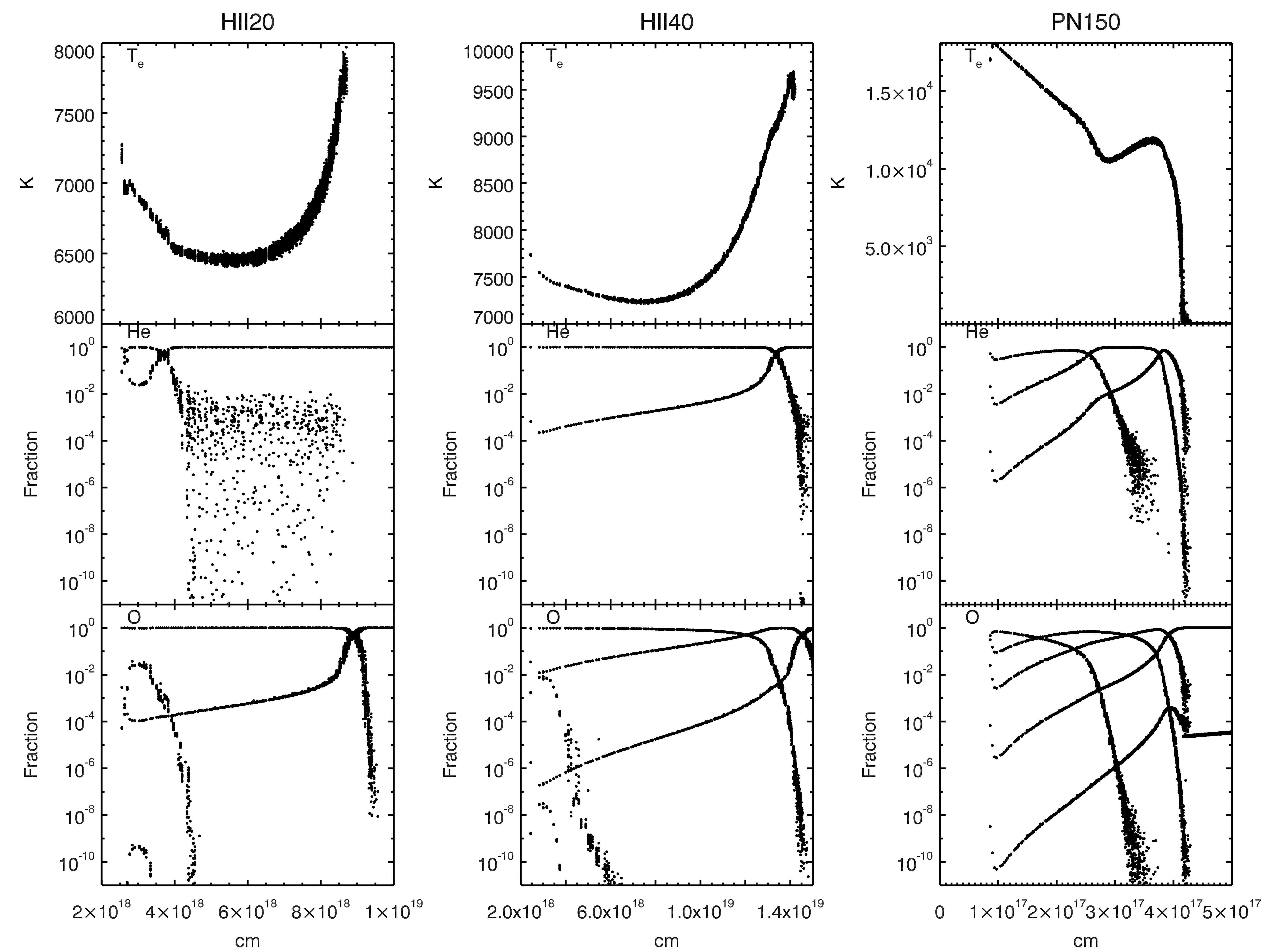}
  \caption{The thermal and ionization structures of HII20, HII40 and PN150 Meudon/Lexington benchmarks. From top to bottom, each row shows the radial profile of the electron temperature and the ionic fraction of He and O.}\label{fig2}
\end{figure*}


We apply our \newcode\ code to the Lexington/Meudon models.
The Lexington/Meudon standard models are a series of artificial cases designed by modellers at the workshops in Meudon, France \citep{Pequignot-1986} and Lexington, Kentucky, USA \citep{Ferland-1995}, for the purpose of testing the capability of each code in modelling low-temperature, high-temperature \hiireg\ regions and planetary nebula. 

We run three Lexington/Meudon benchmarks, HII20, HII40 and PN150.
All the three benchmarks are run in a low-resolution mode with $33^3$ cells and a high-resolution mode with $55^3$ cells.
All the models are run on the computer with a 2.9-3.3GHz CPU.
The HII20 model represents the physical conditions in low-temperature \hiireg\ regions with 20000~K and the HII40 model represents high-temperature \hiireg\ regions with 40000~K.
The PN150 model is representative of planetary nebulae which have hard ionization fields compared to normal \hiireg\ regions.
The details of each model are listed in Table~\ref{tab:benchmark}.

Figure~\ref{fig2} presents the radial profiles of electron temperature and the ionic fraction predicted by \newcode.
We reproduce the tight temperature and ionic fraction profiles.
The $\rm He^{+}$ fraction scatters largely beyond the $\rm He^{+}$ ionizing front in the HII20 model, which is from the residual ionization caused by charge exchange reactions between hydrogen and helium. 
However, the value of $\rm He^{+}$ fraction is below $\rm 1$ per cent, which has no impact on the temperature, electron density and emission-line fluxes.

Predicted emission-line fluxes are listed in Table~\ref{tab:HII20} to Table~\ref{tab:PN150}. 
We also show the results given by other photoionization codes and calculate median fluxes for each emission-line as a reference.
These reference photoionization codes are {\sc cloudy} \citep[][hereafter GF]{Ferland-1995}, \mappings\ \citep[][hereafter RS]{Sutherland-1993}, {\sc mocassin3d} \citep[][hereafter BE]{Ercolano-2003}, {\sc nebula} \citep[][hereafter RR]{Rubin-1991} and P. Harrington's code (hereafter PH).
{\sc cloudy} and {\mappings} are two popular one-dimensional photoionization codes using the $outward$-$only$ approximation on the treatment of diffuse radiation field.
The RR and PH codes are chosen because they are the only two codes solving the diffuse radiation transfer by iterative calculations.
{\sc mocassin3d} is a Monte Carlo photoionization code developed with a different treatment of microphysics. 
Compared with {\sc mocassin3d}, \newcode\ has a more comprehensive consideration of the ionization and recombination processes in the ISM.

In the HII20 and PN150 models, all emission-line fluxes predicted by \newcode\ are within $\rm 2\sigma$ deviation to the reference fluxes.
In the low-resolution mode HII40 model, {[O~\sc i]}$\lambda6300+6363$ and {[S~\sc ii]}$\lambda6716+6731$ are $\rm 2-3\sigma$ of the reference emission-line fluxes. 
In the high-resolution mode HII40 model, the \hb\ fluxes are $\rm 2-3\sigma$ of reference emission-line fluxes.

\section{The Effect of Spatial Resolution on Emission-Line Fluxes}

We compare the emission-line fluxes between the low-resolution and high-resolution \newcode\ models.
Most emission-line predictions are consistent between the two resolution modes with less than 10 per cent difference in flux. 
In contrast, the differences in emission-line fluxes are greater than 10 per cent for  {[O~\sc i]}$\lambda6300+6363$, {[O~\sc ii]}$\lambda7230+7330$, {[S~\sc ii]}$\lambda6716+6731$ and {[S~\sc ii]}$\lambda4068+4076$ in the HII40 model, and for {[N~\sc i]}$\lambda5200+5198$, {[O~\sc i]}$\lambda63.1\mu m$, {[O~\sc i]}$\lambda6300+6363$, {[Ne~\sc ii]}$\lambda12.8\mu m$, {Si~\sc ii]}$\lambda2335+$, {[S~\sc ii]}$\lambda6716+6731$ and {[S~\sc ii]}$\lambda4069+4076$ in the PN150 model.
These emission-lines are located at the boundary of photoionized regions, which are sensitive to the spatial resolution of models.

\section{Modeling \hiireg\ Regions with Complex Geometries}\label{sec:models}

Realistic nebulae display diverse geometries.
The geometry of \hiireg\ regions can be roughly classified into blister, bipolar, spherical and irregular \citep{DePree-2005,Deharveng-2015}, which correspond to the blister, bipolar, spherical and fractal \hiireg\ region models respectively.
The spherical \hiireg\ region can be modeled by previous one-dimensional photoionization codes.
We apply \newcode\ to model three fiducial complex geometries of real nebulae, which are blister, bipolar and irregular \hiireg\ regions, to display the capability of \newcode\ in dealing with complex geometries of nebulae.

\subsection{Blister HII regions}


\begin{figure*}
  \centering
  \includegraphics[width=1\textwidth]{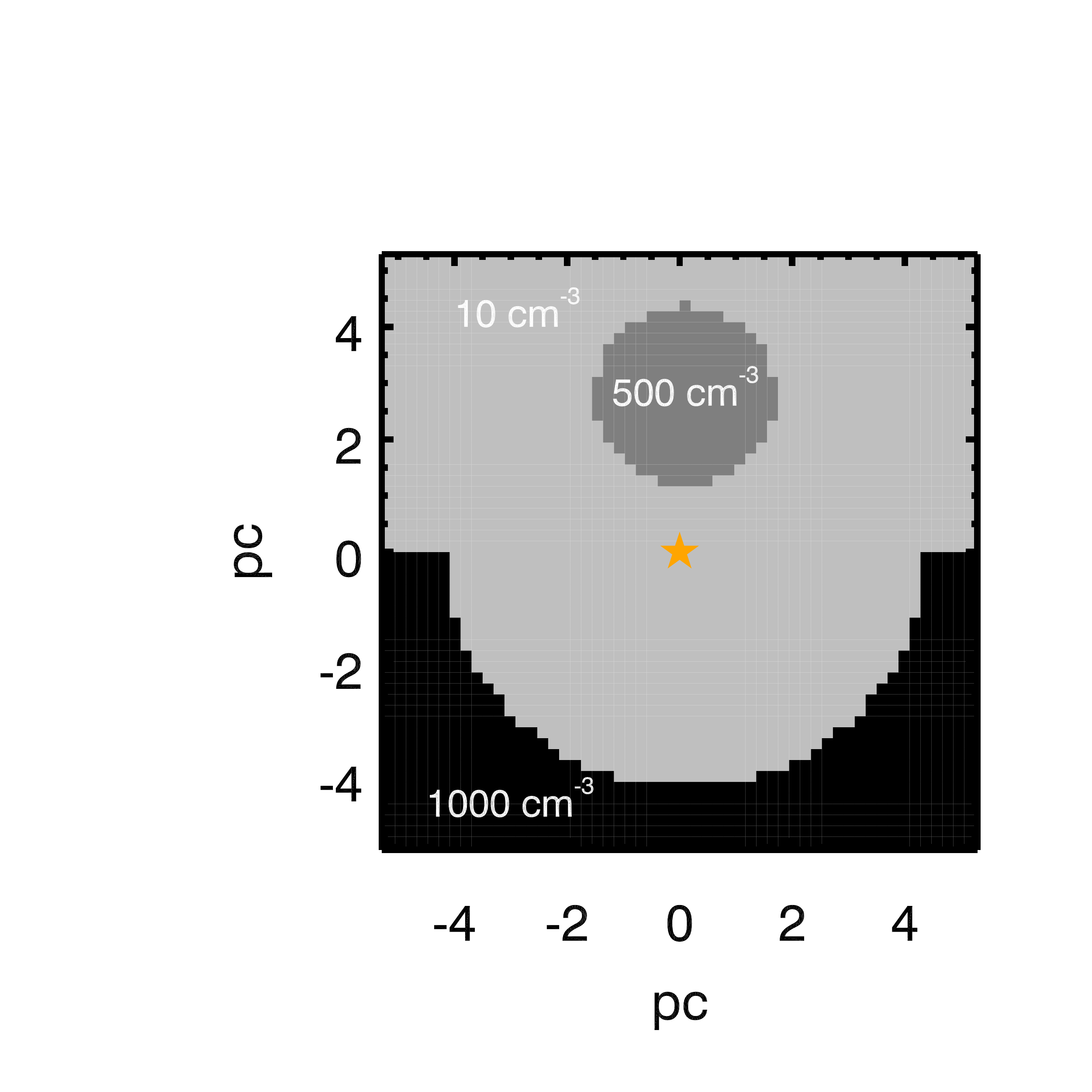}
  \caption{The middle plane (x=0) of the ISM density cube of the blister \hiireg\ region model. The black base is the area with the density of hydrogen of 1000~$\rm cm^{-3}$. The dark grey area shows the intermediate density region with the density of 500~$\rm cm^{-3}$. The light grey area is the low density area with the density of 10~$\rm cm^{-3}$.The central orange star indicates the position of the ionizing source.}\label{fig4}
\end{figure*}

\begin{figure*}
  \centering
  \includegraphics[width=1\textwidth]{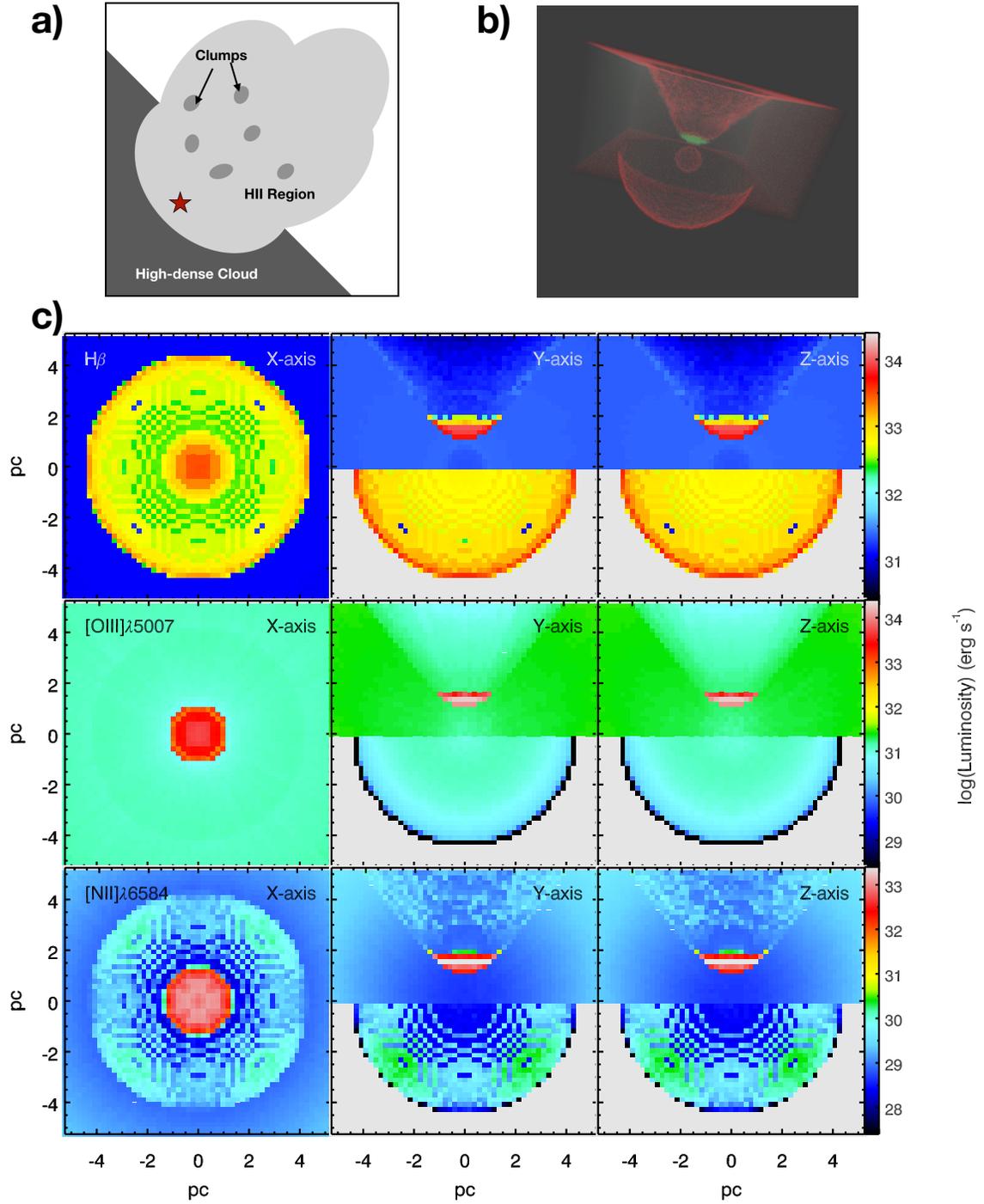}
  \caption{{\bf{a)}} Schematic figure of a blister \hiireg\ region model. A blister \hiireg\ region consists of three major components: a high density cloud base, a intermediate density clumps and the low density ionized gas. {\bf{b)}} Three-dimensional visualization of the modeled blister \hiireg\ region. {\bf{c)}} Distribution of the emission-line luminosity integrated along the x-axis (left), y-axis (middle) and the z-axis (right). We present the distributions of the \hb , \oiii\ and \nii\ emission-lines.}\label{fig5}
\end{figure*}

\begin{figure*}
  \centering
  \includegraphics[width=1\textwidth]{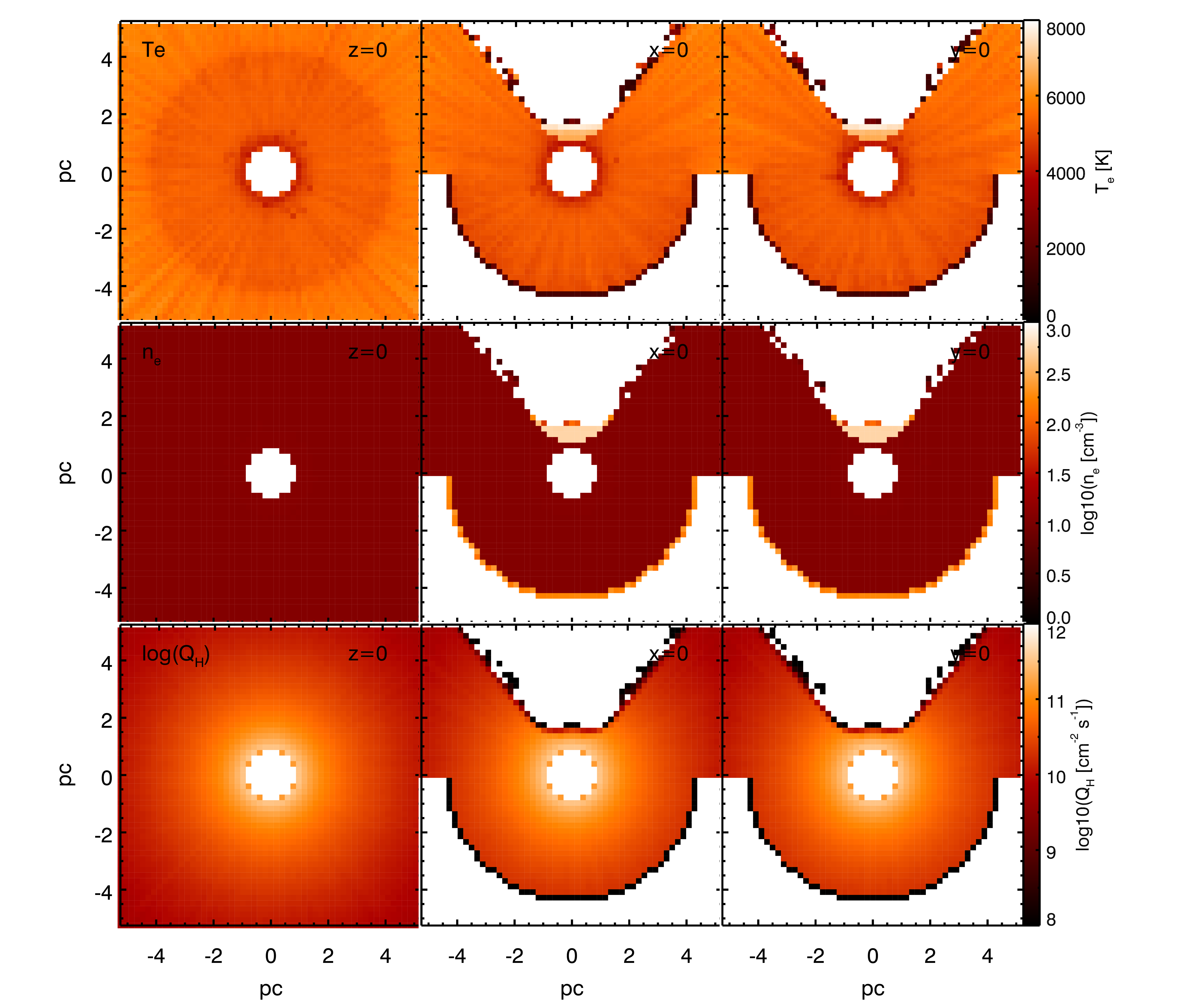}
  \caption{Slices of distributions of the electron temperature, the electron density and the H-ionizing photon flux. We show the cut at z=0 (left), x=0 (middle) and y=0 (right). }\label{fig6}
\end{figure*}

\begin{figure*}
  \centering
  \includegraphics[width=1\textwidth]{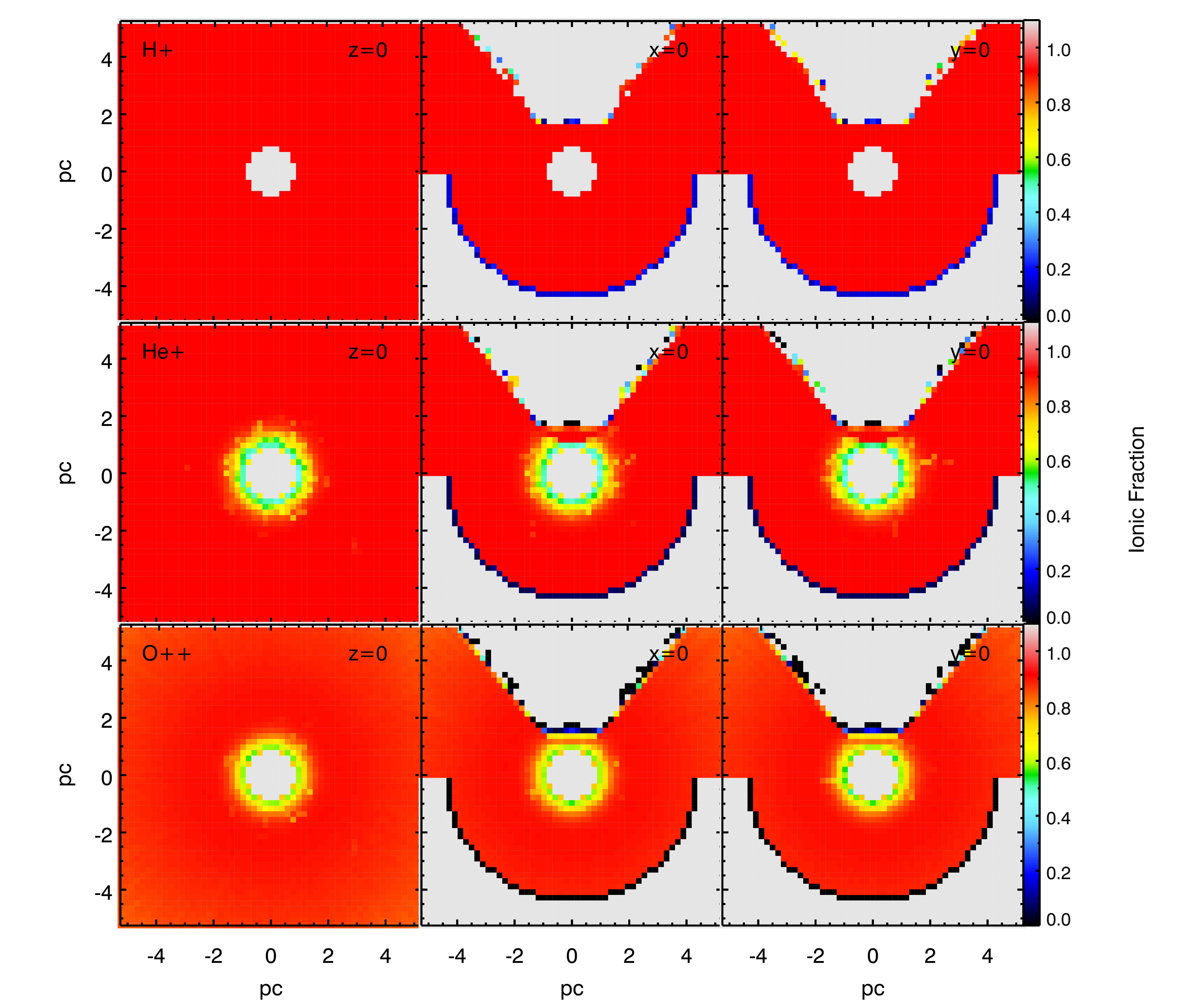}
  \caption{Slices of distributions of $\rm H^{+}$, $\rm He^{+}$ and $\rm O^{++}$. We show the cut at z=0 (left), x=0 (middle) and y=0 (right). }\label{fig7}
\end{figure*}


The blister model is a simplified \hiireg\ region model, where half of the \hiireg\ region is embedded in the dense molecular cloud and half of the \hiireg\ region is expelling into the low-dense clumpy ISM \citep{Tenorio-Tagle-1979,Duronea-2012,Panwar-2020}.

Figure~\ref{fig4} shows the input neutral hydrogen density distribution through the middle panel (x=0) of the blister model.
The ISM density distribution consists of three parts: a low density component with $n_H=10~\rm cm^{-3}$, an intermediate density cloud with $n_H=500~\rm cm^{-3}$ and a high density zone with $n_H=1000~\rm cm^{-3}$. 
The value of $\rm 10~cm^{-3}$ is consistent with the average density of the hydrogen atom in interstellar space \citep{Brinks-1990}.
The intermediate density cloud is designed to reproduce the clumpiness of the ISM.
The high density zone is designed as a ``bowl-like'' shape, consistent with the scenario that blister \hiireg\ regions are caused by the stars formed at the edge of the cloud \citep{Gendelev-2012}.

We select a blackbody with temperature of 40000~K as the simple ionizing source.
The ionizing source is placed at the center of the simulated domain, with a total luminosity $L_{tot}=3.1\times10^{39}~\rm erg~s^{-1}$.
The inner radius of nebula is $R_{in}=3\times10^{18}~\rm cm$.
We adopt the solar abundance set given by \cite{Asplund-2009} (hereafter AS09) as the ISM chemical abundance.

Figure~\ref{fig5} presents the three-dimensional shape of the blister \hiireg\ region model. 
The blister model produces an azimuthally-symmetric nebula consisting of two components. 
The upper component is the emission from a partially ionized clump and the lower component is the emission from the high density ISM. 

\newcode\ successfully reproduces the ionization cone and the diffused ionized gas behind the intermediate gas clump.
The high density and low density gaseous components represent dense natal molecular clouds and the ionized bubble created by ionization front, stellar wind and photon pressure \citep{Gendelev-2012}.
The intermediate density gaseous component represents clumps in the ionized bubble.

We further display the line-of-sight emission-line maps of the blister model in Figure~\ref{fig5} to mimic imaging observations.
Similar to spherical nebula models, the \hiireg\ region observed along the x-axis has a spherical geometry.
However, the integrated emission-line map shows two separate components in views along the y-axis and z-axis.

We trace the distribution of the \hb , \nii\ and \oiii\ diagnostic lines given their importance in separating ionization sources and in measuring metallicity and ionization parameters in galaxies and \hiireg\ regions.
The \hb\ and \nii\ spatially coexist within the blister \hiireg\ region, where the majority of luminosity are from the high density ISM.
The lower component contributes 80 per cent of the total \hb\ luminosity and 87 per cent of the total \nii\ luminosity. 
The \oiii\ emission is predominantly from the intermediate density clump which is close to the ionizing source.
The clump is partially ionized but contributes 96 per cent of the total \oiii\ luminosity.

Figure~\ref{fig6} shows the slices of electron temperature, electron density and H-ionizing photon flux across planes of x=0, y=0 and z=0.
The electron temperature, the electron density and the H-ionizing photon flux are uniformly distributed within the main body of the nebula.
The main body of the nebula has an average electron temperature of 4000~K and an average electron density of 10~$\rm cm^{-3}$.
The intermediate density cloud has a hotter average electron temperature of 6000~K and a higher average electron density of 450~$\rm cm^{-3}$ than the remaining part of the nebula.
The average electron temperature in the diffuse ionized gas is 1000~K cooler than the main body of the nebula.  
The average electron density in the diffuse ionized gas is 10~$\rm cm^{-3}$ similar to the main body of the nebula.

Figure~\ref{fig7} shows the slices of ionic fractions of $\rm H^{+}$, $\rm He^{+}$ and $\rm O^{++}$ across planes of x=0, y=0 and z=0.
The $\rm H^{+}$ and $\rm He^{+}$ are dominant ionic species within nebulae and the $\rm O^{++}$ is the major coolant species.
The ionic fractions of $\rm H^{+}$, $\rm He^{+}$ and $\rm O^{++}$ are uniformly distributed across the nebula.

\subsection{Bipolar HII regions}

\begin{figure*}
  \centering
  \includegraphics[width=1\textwidth]{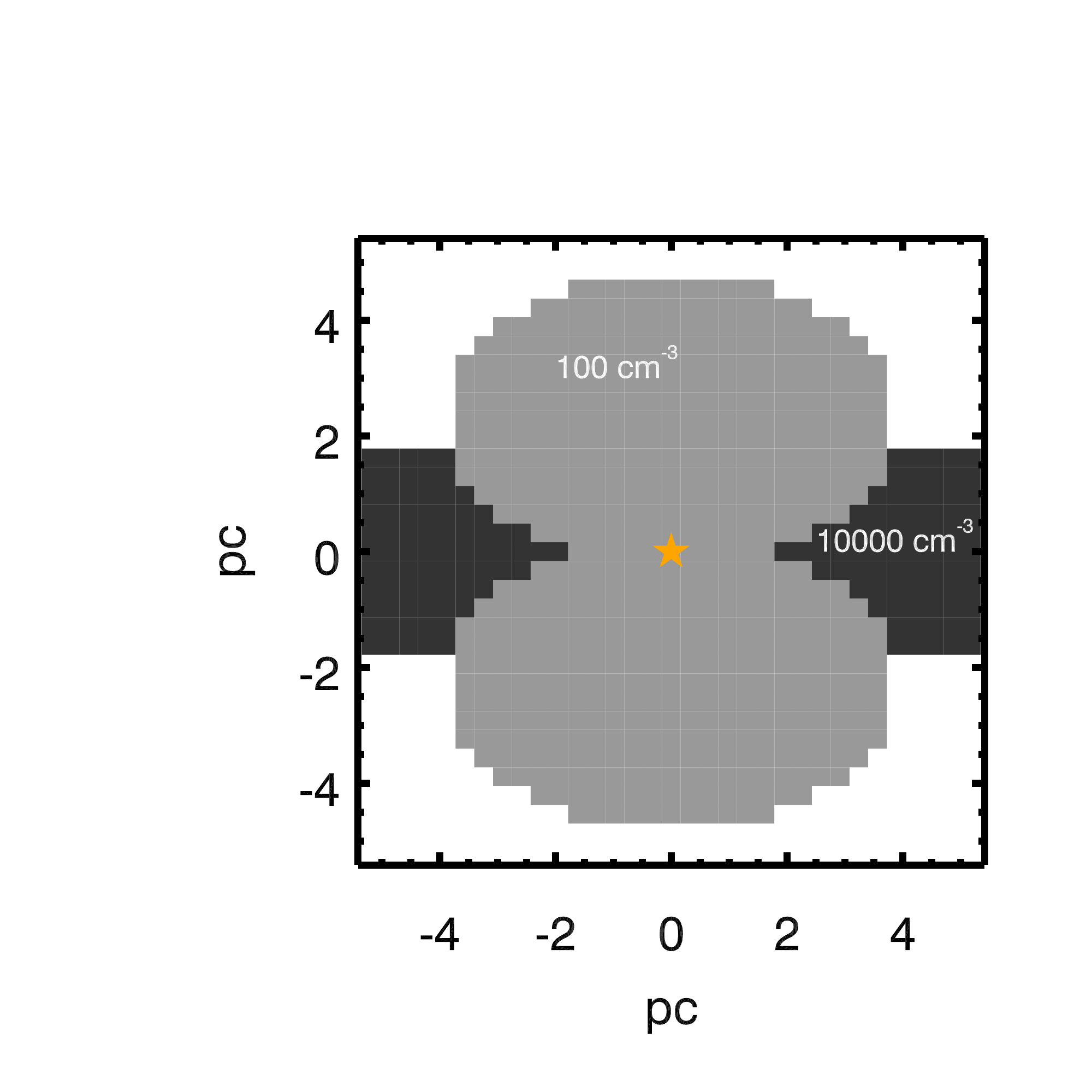}
  \caption{The middle plane (x=0) of the ISM density cube of the bipolar \hiireg\ region model. The black area represents the sheet-like high density cloud with the density of hydrogen of 10000~$\rm cm^{-3}$. The grey area shows the ionized gas with the density of 100~$\rm cm^{-3}$.The central orange star indicates the position of the ionizing source.}\label{bipolar_input}
\end{figure*}

\begin{figure*}
  \centering
  \includegraphics[width=1\textwidth]{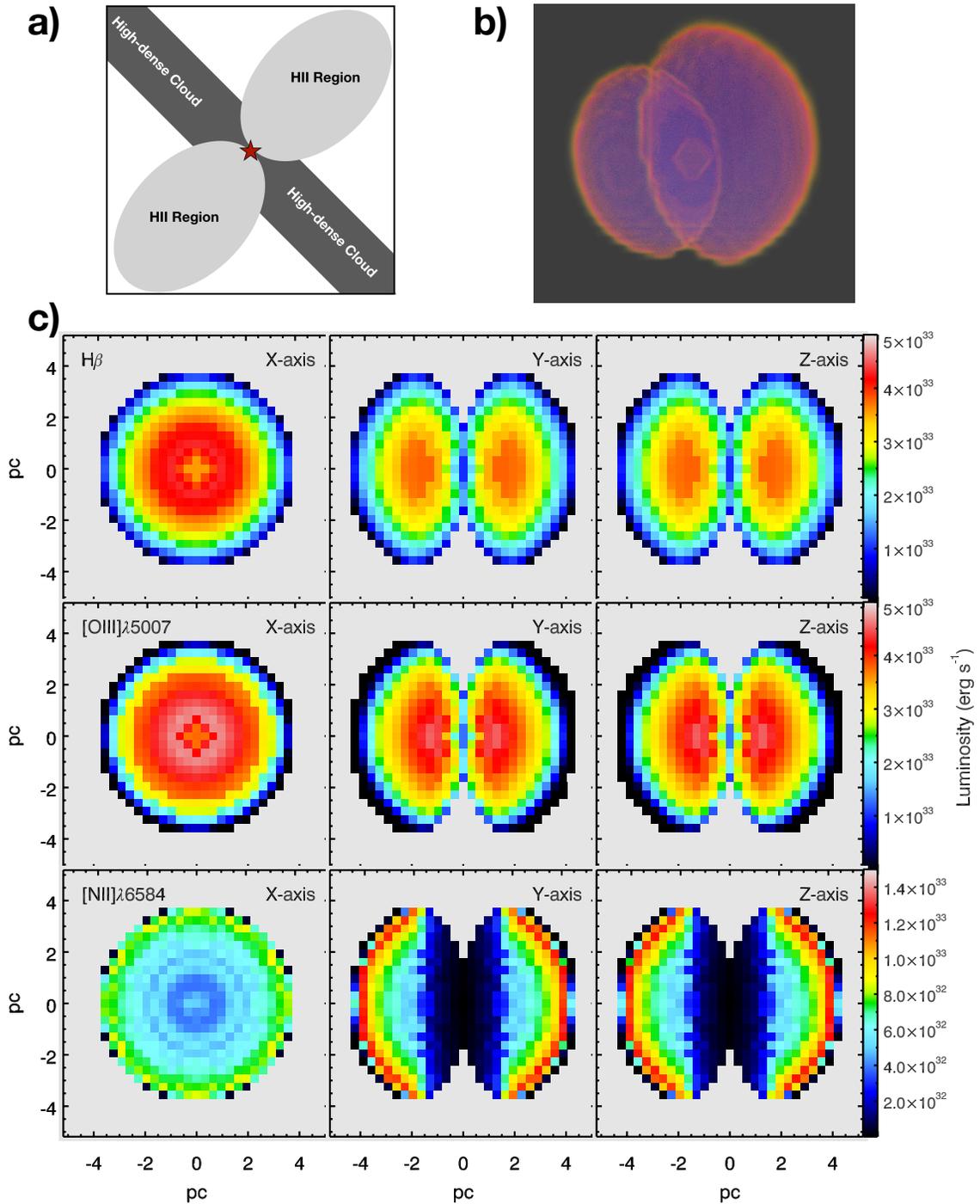}
  \caption{{\bf{a)}} Schematic figure of a bipolar \hiireg\ region model. A bipolar \hiireg\ region consists of two major components: a high-dense sheet-like cloud, the low-dense ionized gas. {\bf{b)}} Three-dimensional visualization of the modeled bipolar \hiireg\ region. {\bf{c)}} Distribution of the emission-line luminosity integrated along the x-axis (left), y-axis (middle) and the z-axis (right). We present the distributions of the \hb , \oiii\ and \nii\ emission-lines.}\label{fig8}
\end{figure*}

\begin{figure*}
  \centering
  \includegraphics[width=1\textwidth]{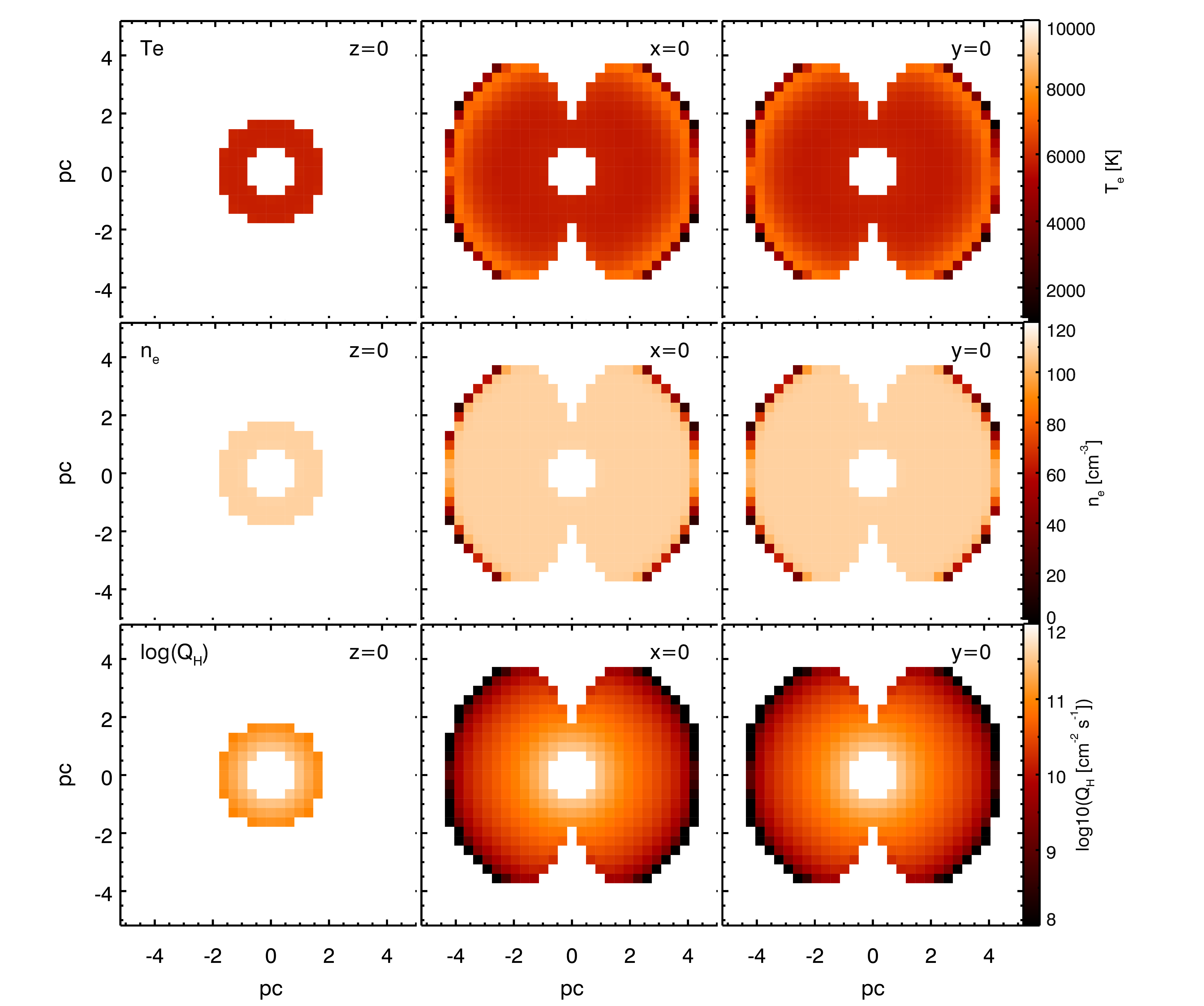}
  \caption{Slices of distributions of the electron temperature, the electron density and the H-ionizing photon flux. We show the cut at z=0 (left), x=0 (middle) and y=0 (right). }\label{fig9}
\end{figure*}

\begin{figure*}
  \centering
  \includegraphics[width=1\textwidth]{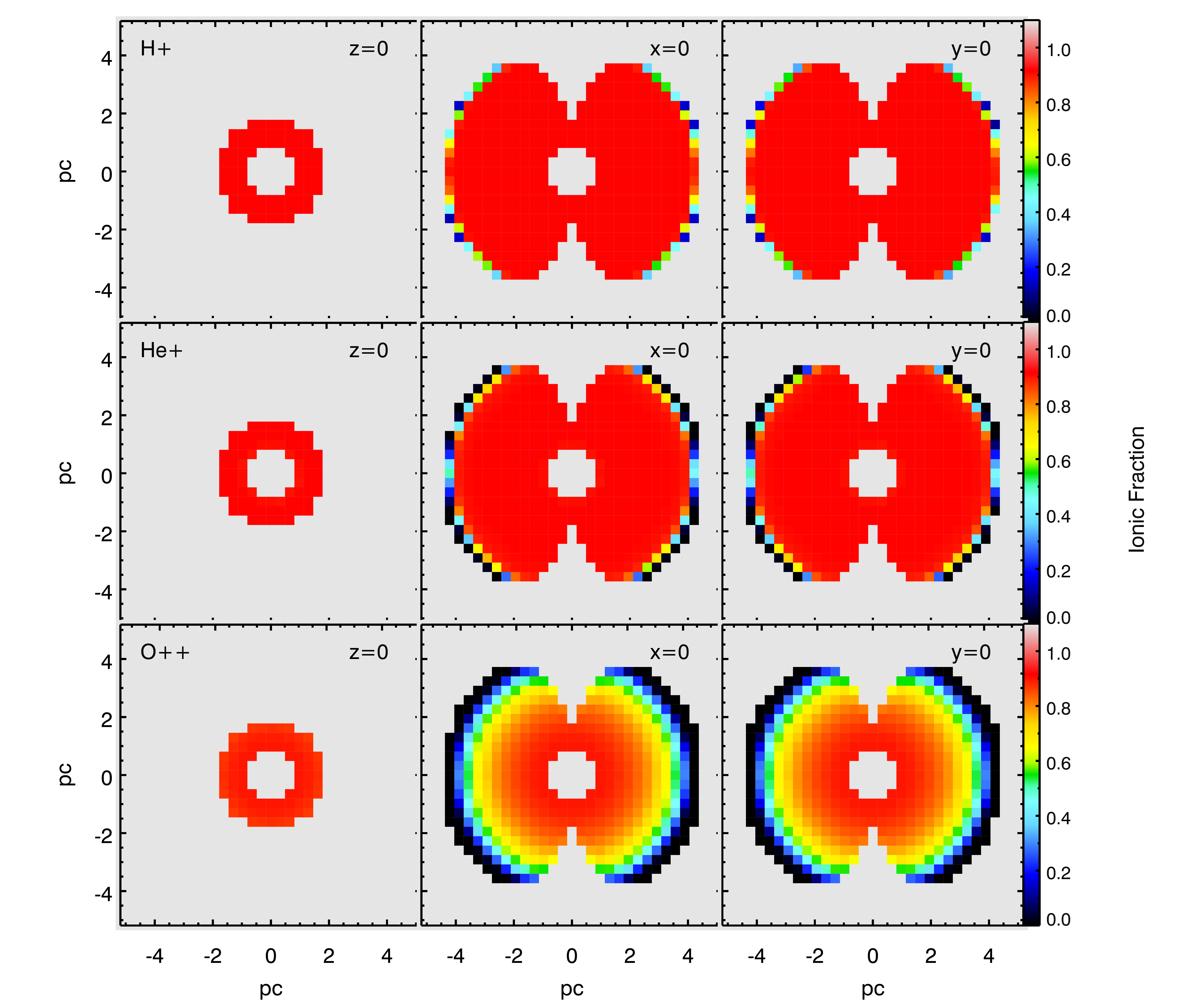}
  \caption{Slices of distributions of $\rm H^{+}$, $\rm He^{+}$ and $\rm O^{++}$. We show the cut at z=0 (left), x=0 (middle) and y=0 (right). }\label{fig10}
\end{figure*}


The bipolar \hiireg\ region model is composed of two ionized lobes located perpendicularly to a dense molecular cloud \citep{Samal-2018}.
Three-dimensional simulations suggest that the bipolar structure is the consequence of a star evolving in a sheet-like molecular cloud \citep{Bodenheimer-1979,Wareing-2017}.

Figure~\ref{bipolar_input} shows the initial condition of neutral hydrogen density distribution through the middle panel (x=0) of the bipolar \hiireg\ region model.
The ISM density distribution consists two lobes with $n_H=100~\rm cm^{-3}$ oriented perpendicular to a sheet-like high density cloud with $n_H=10000~\rm cm^{-3}$. 
The density of $\rm 100~cm^{-3}$ within two lobes is selected to match with the typical density of local nebula in star-forming galaxies \citep{Kewley-2001,Levesque-2010}.
The density of the high density cloud is selected based on data that indicate that the sheet-like clouds have density of $\rm 10^4~cm^{-3}$ \citep{Deharveng-2015}.

The central ionizing source is selected as a blackbody with temperature of 40000~K and a total luminosity $L_{tot}=3.1\times10^{39}~\rm erg~s^{-1}$.
The inner radius of nebula is $R_{in}=3\times10^{18}~\rm cm$.
The ISM chemical abundance is the AS09 solar abundance.

Figure~\ref{fig8} presents the three-dimensional shape of the bipolar \hiireg\ region model. 
We present the line-of-sight distribution of the integrated \hb , \oiii\ and \nii\ emission-lines.
The \hb\ and \oiii\ emissions are filled in the bipolar lobes appearing as two bubble structures.
By contrast, the \nii\ emissions are mainly located on the surface of the bipolar lobes, appearing to be a shell-like morphology.

The bubble structures in the bipolar model exist in most \hiireg\ regions \citep{Churchwell-2007,Deharveng-2015}.
Bipolar structures may appear as ring-like structures if the nebulae are observed in particular viewing angles \citep{Anderson-2011}. 
The 3D visualization of our bipolar model presents ring-like structures similar to real observations.

Figure~\ref{fig9} shows the slices of electron temperature, electron density and H-ionizing photon flux across planes of x=0, y=0 and z=0. 
The radial profiles of electron temperature, electron density and H-ionizing photon flux in the bipolar nebula are similar to the profiles of the spherical model, where the electron temperature, the electron density and the H-ionizing photon flux have flat gradients within the nebula.
The average electron density is 108~$\rm cm^{-3}$ and the average electron temperature is 6000~K.
The electron density is reduced and the electron temperature becomes cooler at the edge of the nebula because the ionizing photons are totally absorbed.

Figure~\ref{fig10} shows the slices of ionic fractions of $\rm H^{+}$, $\rm He^{+}$ and $\rm O^{++}$ across planes of x=0, y=0 and z=0.
The ionic fractions of $\rm H^{+}$ and $\rm He^{+}$ are uniformly distributed in the nebula.
In contrast, the ionic fraction of $\rm O^{++}$ decreases along with the radius of the nebula.

\subsection{Fractal Geometry HII regions}

\begin{figure*}
  \centering
  \includegraphics[width=.95\textwidth]{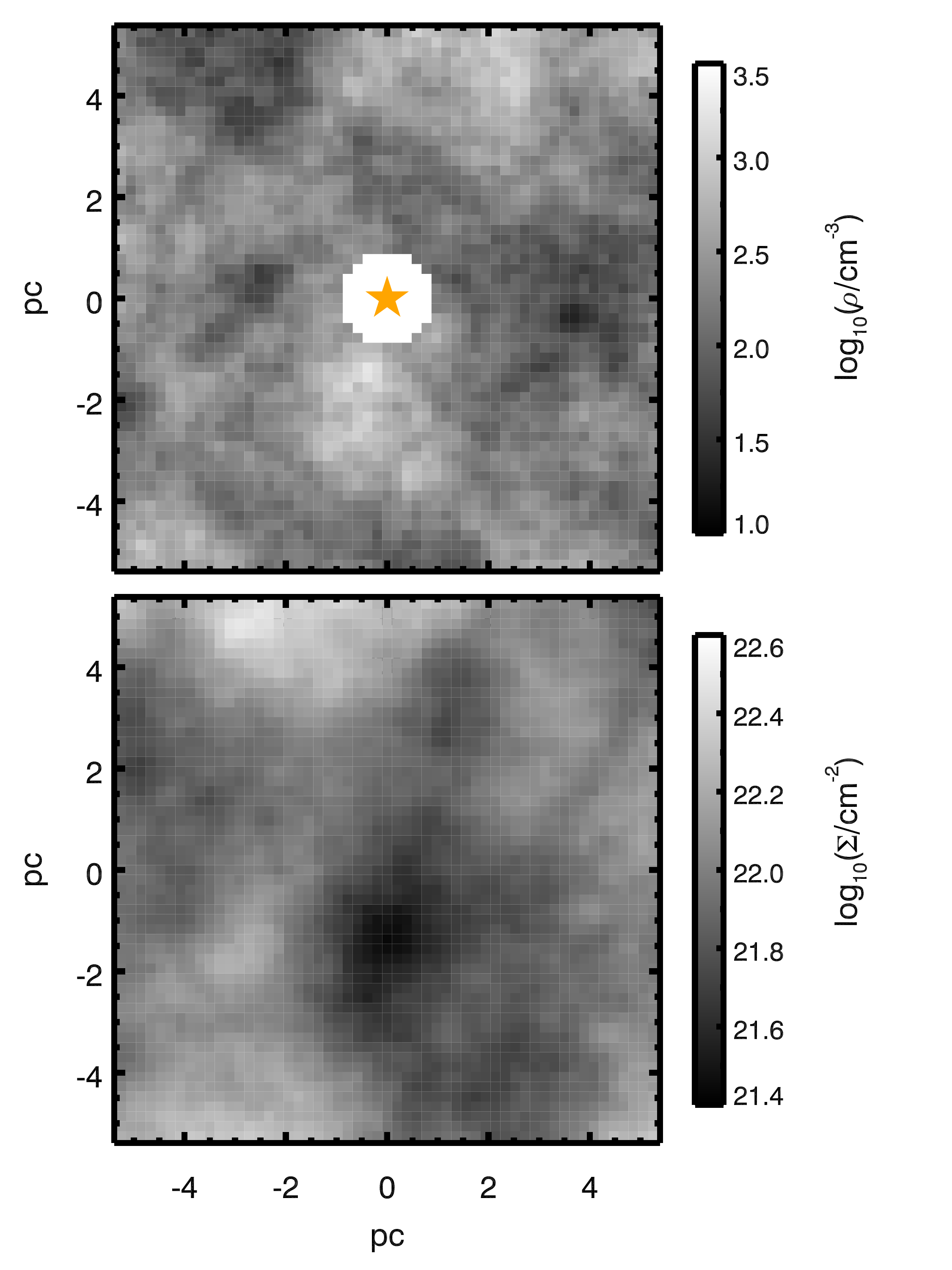}
  \caption{Density distribution of the fractal ISM set up for the fractal \hiireg\ region model. The top panel is the middle plane of the input cube of number density, $\rho$. The central cavity corresponds to an inner radius $R_{in}=3\times10^{18}~\rm cm$. The orange star indicates the position of the ionizing source. The bottom panel shows the column number density, $\Sigma$, of the input fractal ISM.}\label{fractal_input}
\end{figure*}

\begin{figure*}
  \centering
  \includegraphics[width=1\textwidth]{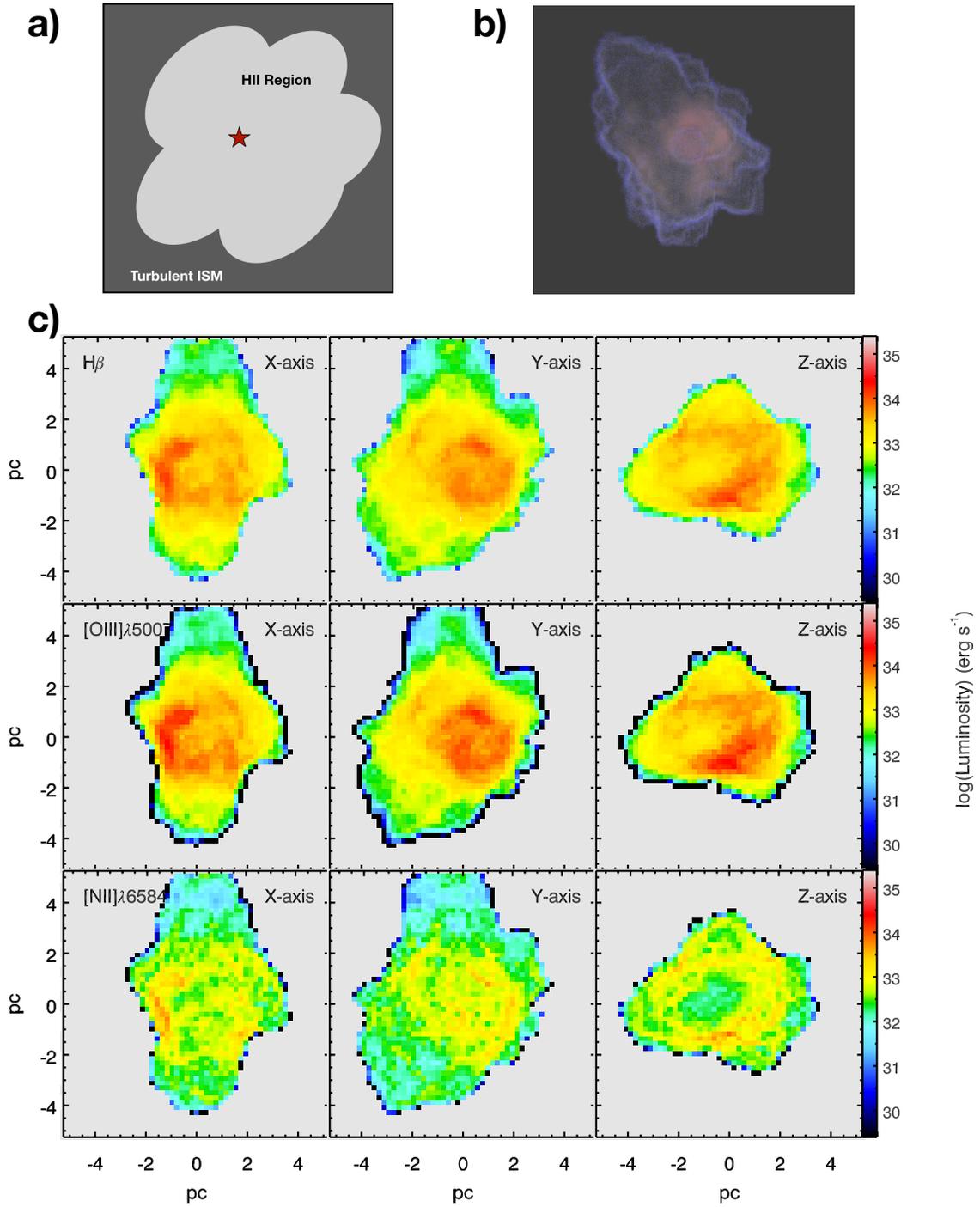}
  \caption{{\bf{a)}} Schematic figure of a fractal \hiireg\ region model. A fractal \hiireg\ region is embedded in the turbulent ISM. {\bf{b)}} Three-dimensional visualization of the modeled fractal \hiireg\ region. {\bf{c)}} Distribution of the emission-line luminosity integrated along the x-axis (left), y-axis (middle) and the z-axis (right). We present the distributions of the \hb , \oiii\ and \nii\ emission-lines.}\label{fig11}
\end{figure*}

\begin{figure*}
  \centering
  \includegraphics[width=1\textwidth]{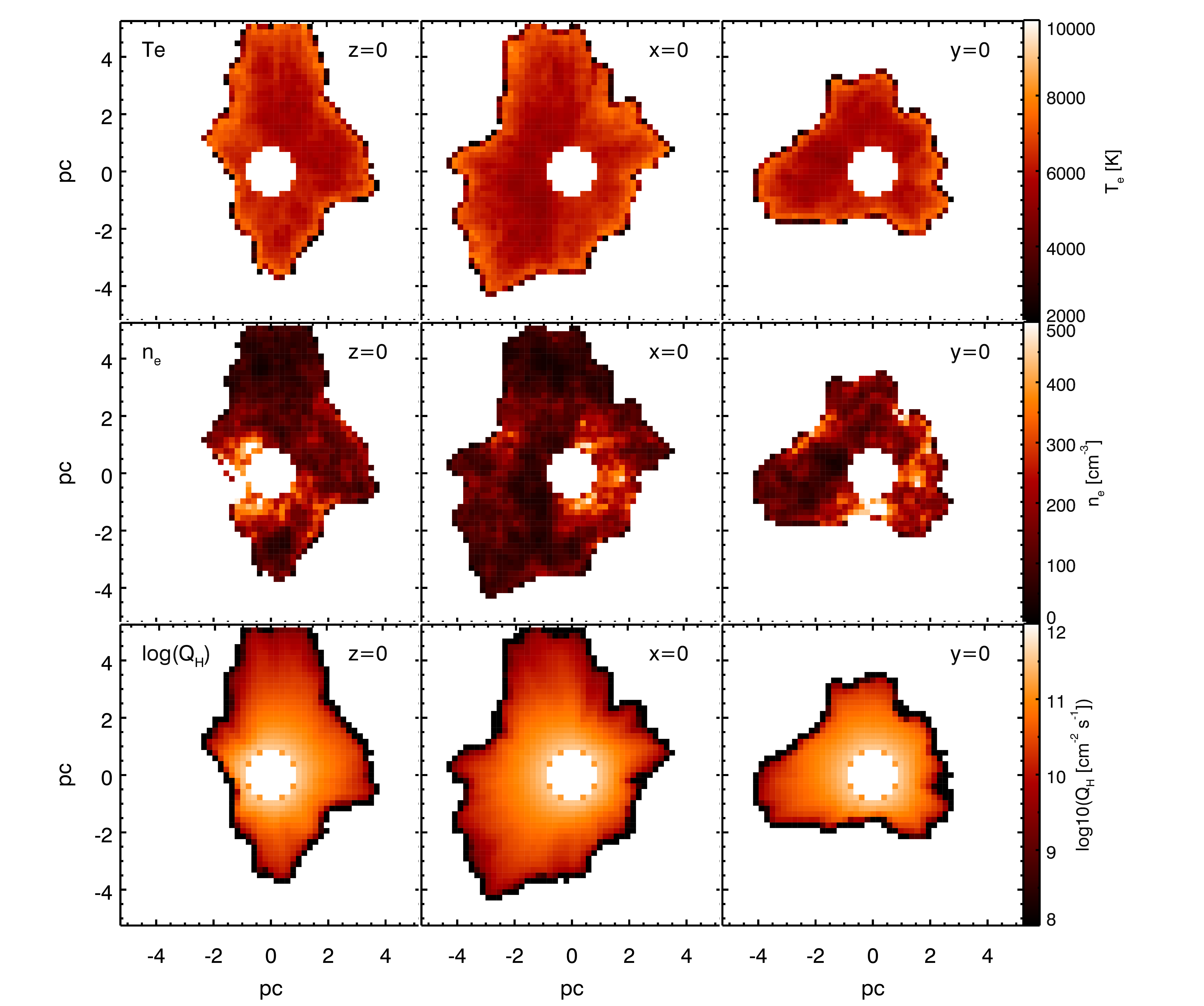}
  \caption{Slices of distributions of the electron temperature, the electron density and the H-ionizing photon flux. We show the cut at z=0 (left), x=0 (middle) and y=0 (right).}\label{fig12}
\end{figure*}

\begin{figure*}
  \centering
  \includegraphics[width=1\textwidth]{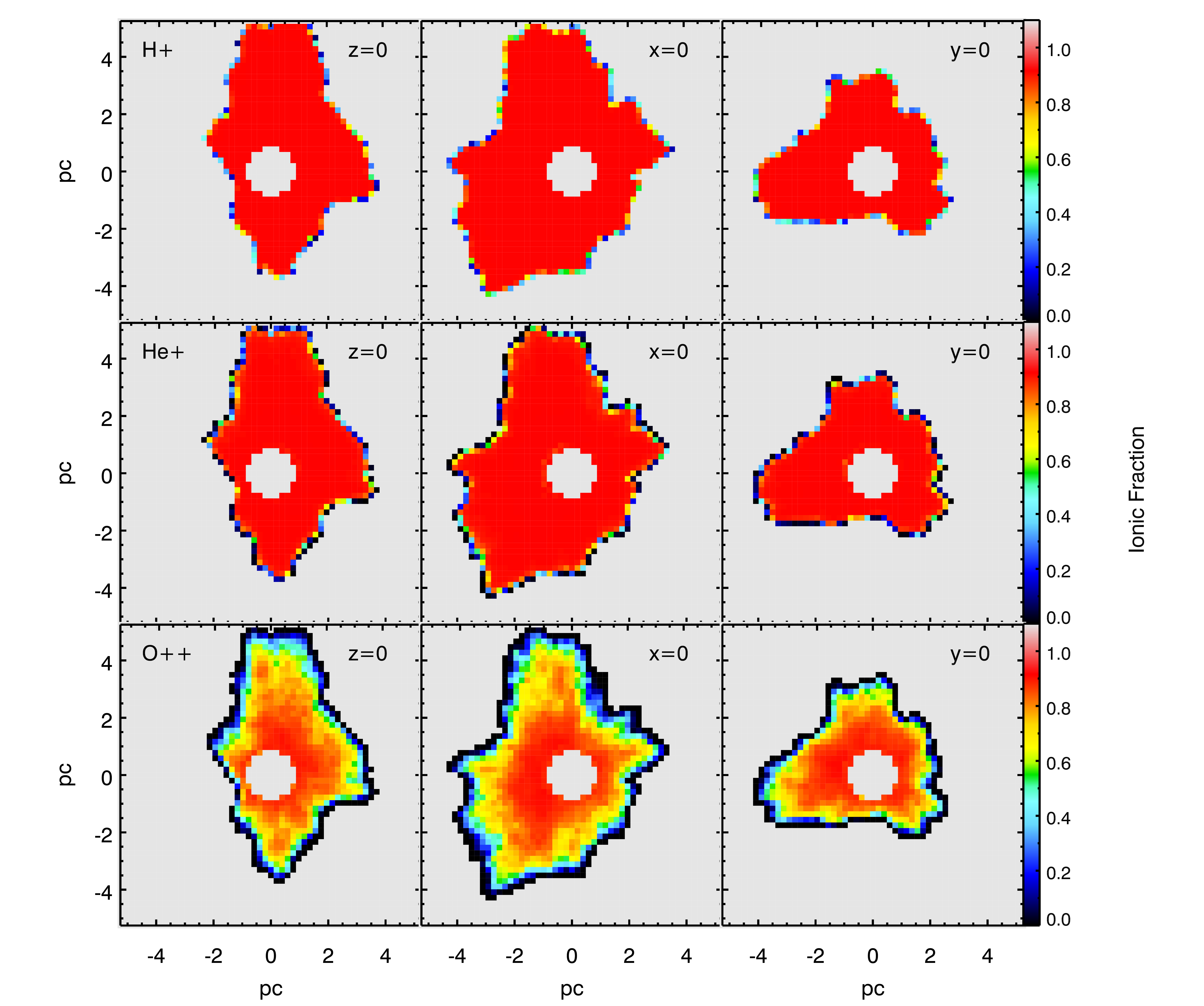}
  \caption{Slices of distributions of $\rm H^{+}$, $\rm He^{+}$ and $\rm O^{++}$. We show the cut at z=0 (left), x=0 (middle) and y=0 (right). }\label{fig13}
\end{figure*}


The fractal \hiireg\ region model represents an \hiireg\ region evolving in the turbulent ISM \citep{Medina-2014} and developing self-similar structures \citep{Zuckerman-1973, Rubin-2011,Arthur-2016, O'Dell-2017}.
Both observations \citep{Wisnioski-2015} and simulations \citep{Pillepich-2019} reveal that the turbulent motion dominates the ISM kinematics especially in high-redshift galaxies.

Figure~\ref{fractal_input} shows the neutral hydrogen density distribution in the fractal model.
Following the Kolmogorov's theory \citep{Kolmogorov-1941}, the power spectrum of the turbulence is in the format of $E(k)\propto k^{-5/3}$, where $k$ is the wavenumber $k\sim1/r$.  
The column density probability distribution function (PDF) of the ISM is designed as the log-normal distribution because of the hierarchical structures of the turbulent ISM \citep{Larson-1981}. 
The mean value of the column density is 100~$\rm cm^{-3}$.
We use AS09 for the ISM chemical abundances.
The central ionizing source is a blackbody with temperature of 40000~K and a total luminosity $L_{tot}=3.1\times10^{39}~\rm erg~s^{-1}$.
The inner radius of nebula is $R_{in}=3\times10^{18}~\rm cm$.

Figure~\ref{fig11} displays the shape of the fractal \hiireg\ region model in three-dimensions. 
The modeled \hiireg\ region has an inhomogeneous geometry which is caused by the fractal density distribution of the ISM.

Theories suggest that a turbulent model best describes the ISM which has a fractal density distribution \citep{Federrath-2009}.
\newcode\ successfully reproduces the fractal \hiireg\ region model with an inhomogeneous density distribution and non-uniform emission-line distributions, which neither spherical models nor plane-parallel models can produce.

We also present the line-of-sight integrated emission-line maps of the fractal model.
The \hiireg\ region morphology changes in different line-of-sight views.
In each line-of-sight direction, the \hb\ and \oiii\ share the similar spatial distribution, which are concentrated around the center of the nebula.
Compared with the \hb\ and \oiii , the \nii\ is less concentrated to the central source because the high local ionization parameter ionizes $\rm N^{+}$ to higher ionization levels. 

Figure~\ref{fig12} shows the slices of electron temperature, electron density and H-ionizing photon flux across planes of x=0, y=0 and z=0. 
The electron temperature and the H-ionizing photon flux are uniformly distributed across the nebula.
In contrast, the electron density distribution shows a large fluctuation, where the high density ISM has the high electron density and the low density ISM has the low electron density.
The average electron density is 150~$\rm cm^{-3}$ with the standard deviation of 90~$\rm cm^{-3}$.

Figure~\ref{fig13} shows the slices of ionic fractions of $\rm H^{+}$, $\rm He^{+}$ and $\rm O^{++}$ across planes of x=0, y=0 and z=0.
The ionic fractions of $\rm H^{+}$ and $\rm He^{+}$ are uniformly distributed across the model.
The ionic fraction of $\rm O^{++}$ decreases as a function of the nebular radius.

\section{Comparison Between Models with Complex Geometries and Spherical Models}

\begin{figure*}
  \centering
  \includegraphics[width=1\textwidth]{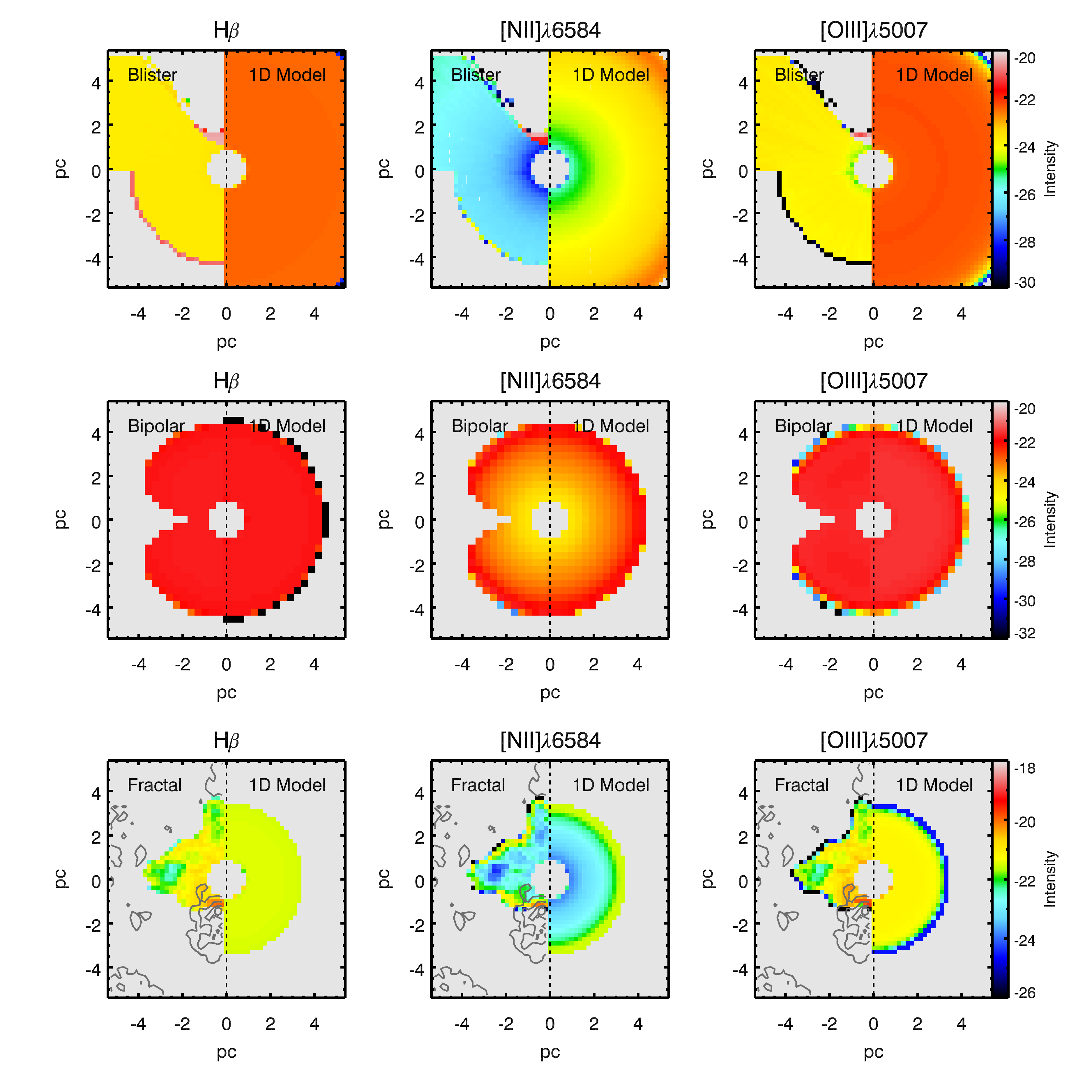}
  \caption{Comparison of the emission-line distribution between the nebular models with complex geometries and the corresponding spherical models. In each panel, the left side shows the cut of the emission-line distribution of the model with complex geometry. The right side shows the cut of the emission-line distribution of the spherical model.}\label{fig14}
\end{figure*}

\begin{figure*}
  \centering
  \includegraphics[width=1\textwidth]{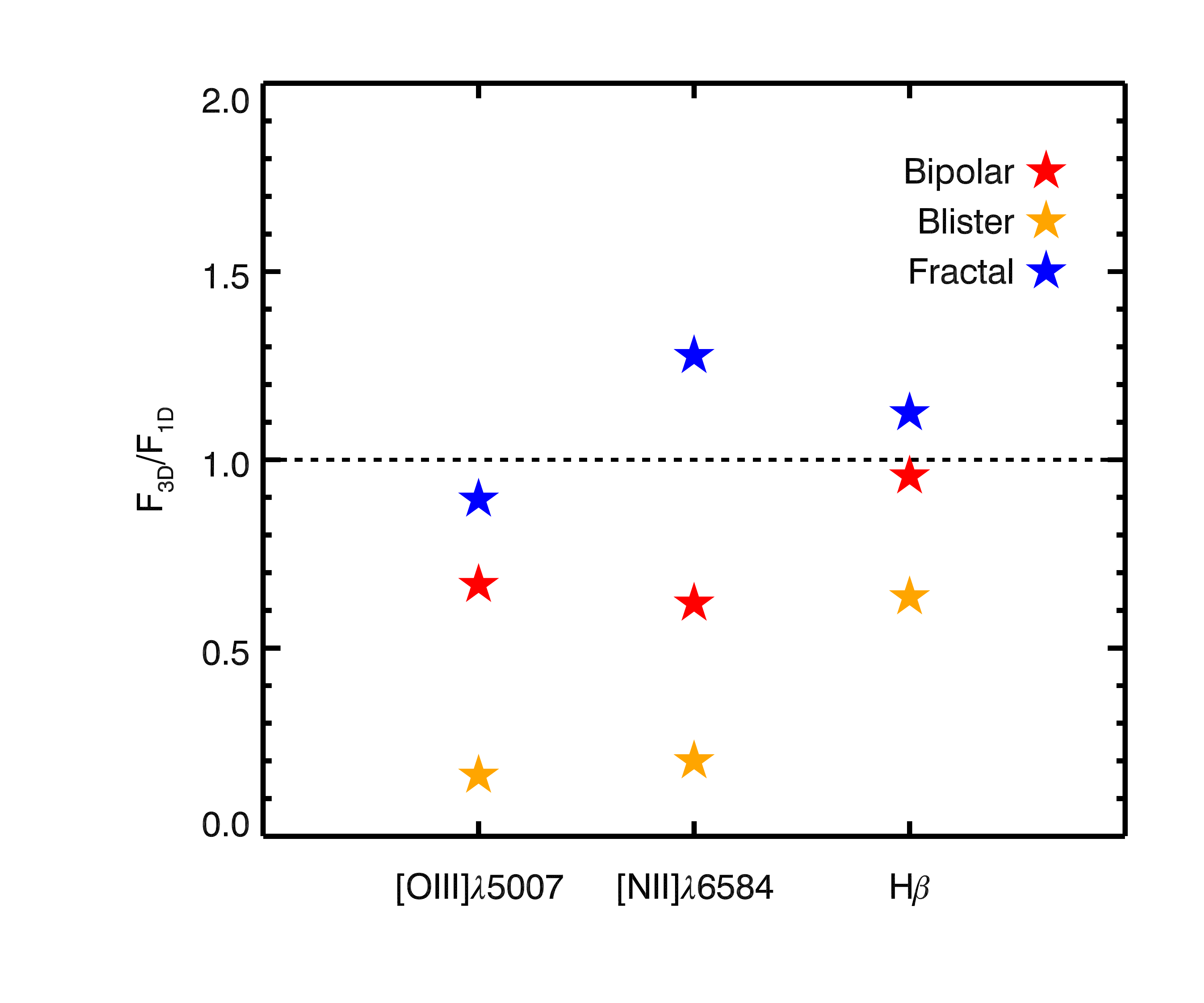}
  \caption{Comparison of integrated emission-line fluxes (left) and emission-line ratios (right) between the models with complex geometries and the spherical models. The horizon dashed line indicates that the integrated fluxes or emission-line ratios from models with complex geometry and from spherical models are the same. }\label{fig15}
\end{figure*}

We create a corresponding spherical model for the blister, bipolar and fractal nebula models respectively.
Except for the geometry, the spherical model has the same initial conditions as the geometric models, including the ionizing source, the ISM abundance, the total mass of nebula and the average ISM density.
The corresponding spherical models to the blister and bipolar \hiireg\ regions are radiation-bounded because the radiation field is absorbed before reaching the edge of ISM.
The corresponding spherical model to the fractal \hiireg\ region is density-bounded, because the ISM density inhomogeneity may shorten the radiation field. 

Figure~\ref{fig14} presents the comparison of emission-line distributions between the geometric models and the spherical models.
The spherical models have the obvious stratification of emission-line distributions, where the \nii\ lines are located at the boundary of the nebula and the \oiii\ lines are mainly from the center of the nebula.
By contrast, the density fluctuations in nebulae with complex geometries smooth the stratification of internal distributions of emission-lines.

Figure~\ref{fig15} further shows the changes of integrated emission-line fluxes between the models with complex geometries and the spherical models.
Different nebular geometries change the integrated emission-line fluxes in different directions.
Compared to the spherical model, the ``Blister'' geometry reduces the integrated fluxes of \oiii\ and \nii\ by around 80\% and the integrated flux of \hb\ by around 35\%.
In the bipolar model, the bipolar structure reduces the integrated fluxes of \oiii\ and \nii\ by 30-40\%. 
However, the bipolar model has similar integrated flux of \hb\ compared to the spherical model.
The fractal geometry affects integrated fluxes in the opposite direction.
Compared to the spherical model, the fractal geometry increases the integrated flux of \nii\ by around 30\%\ and the integrated flux of \hb\ by around 10\%. 
The fractal geometry slightly reduces the integrated flux of \oiii\ by 10\%\ compared to the spherical model.

\section{Discussion}\label{sec:discussion}

\subsection{Electron Temperature Structure of HII regions}

Realistic \hiireg\ regions have significant temperature fluctuations within the nebulae \citep{Rubin-2011,Peimbert-2019}.
Detailed observations of $\theta^{1}$ Ori C in the Orion Nebula show that the electron temperature has an increasing trend with the distance from the ionizing star.
In current constant density photoionization models, the electron temperature becomes hotter towards the edge of \hiireg\ regions when the metallicity of nebulae $\rm 12+log(O/H)\ge8.5$ \citep{Kewley-2019}.

The mechanisms leading to the electron temperature fluctuations are still under debate.
Multiple mechanisms have been proposed, including the turbulence and shocks in the ISM \citep{Peimbert-1991,O'Dell-2015}, and the stellar winds produced by central ionizing sources \citep{Gonzalez-Delgado-1994}

Here we propose that the nebular geometry is one, but not the only cause of the electron temperature fluctuation.
The degree of electron temperature fluctuation increases with the complexity of nebular geometry.
The electron temperature in the blister and bipolar \hiireg\ region models has a flat gradient because the hydrogen density is uniformly distributed within ionized bubbles. 
In the blister \hiireg\ region model, the intermediate density clump is hotter than the average electron temperature of the nebula because the clump is denser than the average density of the nebula.
The fractal \hiireg\ region model has a more significant electron temperature fluctuation than the blister and bipolar \hiireg\ region models, because the self-hierarchical structure in the fractal model is the most complex geometry among the blister, bipolar and fractal models. 

Nebula models allowing the electron temperature fluctuation offers great promise in interpreting emission-line spectra.
The spectra of nearby or distant galaxies captured by fixed-size apertures combine the light from an ensemble of \hiireg\ regions with complex electron temperature structures.
The electron temperature fluctuation is a potential cause of the discrepancy between the metallicity determined by recombination lines and those determined with Auroral lines \citep{Peimbert-2019}. 
The turbulent ISM in high-redshift galaxies increase the electron temperature fluctuations of nebulae.
Three-dimensional nebula models with complex electron temperature structures provide more realistic predictions of emission-lines than those models with constant temperature, density or pressure assumptions.

\subsection{Density Structure of HII regions}

Realistic nebulae have significant density fluctuations within local \hiireg\ regions \citep{Perez-2001,Simpson-2004,McLeod-2016,Peimbert-2019}.
Spatially-resolved measurements show negative density gradients in some local \hiireg\ regions \citep{Kurtz-2002,Phillips-2007,Rubin-2011} and flat density gradients in other compact \hiireg\ regions \citep{Garcia-Benito-2010,Ramos-Larios-2010}.
Detailed observations of the Orion Nebula find that the turbulence within the nebula leads to the complex density structures \citep{Arthur-2016, O'Dell-2017,Kewley-2019}.

We present diverse electron density distributions in three fiducial \hiireg\ region models.
The bipolar \hiireg\ region model presents a flat electron density gradient because the density distribution of hydrogen is uniform within the bipolar bubbles.
The blister and fractal \hiireg\ regions present significant electron density fluctuations, which are caused by the inhomogeneity of the ISM density within the nebulae.

\subsection{Emission-Line Predictions}

The \oiii$\lambda5007$, \nii$\lambda6584$ and \hb\ are crucial diagnostic emission-lines for galaxy evolution study.
These three emission-lines have different dependance on the nebular geometry.
The \oiii$\lambda5007$ and \hb\ are produced throughout the entire nebula, and their fluxes are sensitive to the size of nebular volume.
The \nii$\lambda6584$ traces the intermediate-ionization zone of nebula, which is located at the more outer region than the \oiii$\lambda5007$ and \hb .
The flux of \nii$\lambda6584$ is more sensitive to the length of nebular boundary than the fluxes of \oiii$\lambda5007$ and \hb .

Complex geometry affects the emission-line predictions.
The twisted boundary and the irregular shape of the nebula changes the fluxes of \oiii$\lambda5007$, \nii$\lambda6584$ and \hb .
The inhomogeneity of ISM density changes the spatial distributions of emission-lines within nebulae, which cannot be reproduced by nebula models with spherical or plane parallel geometries.
We will investigate the detailed impact of nebular geometry on the optical emission-lines in a
forthcoming paper (Jin et al. in prep).

Oversimplified nebular geometries in emission-line modelings lead to the debate on how the nebulae are bounded.
Spherical and plane parallel nebula models assume that \hiireg\ regions are simply radiation-bounded, in order to reproduce the observed emission-line fluxes in nearby galaxies \citep{Kewley-2001}.
However, \cite{Nakajima-2013} proposed that the \hiireg\ regions in high-redshift galaxies are likely to be density-bounded, where the gas is insufficient to absorb the entire ionizing photons from the star.
Detailed studies of nearby \hiireg\ regions suggest that the radiation-bounded and density-bounded conditions coexist within a single nebula \citep{Pellegrini-2012}.

\newcode\ can produce emission-lines in \hiireg\ region models with complex geometries, allowing a mixture of radiation-bounded and density-bounded cases.
Among the three fiducial nebula models, the bipolar \hiireg\ region model and the fractal \hiireg\ region model are simply radiation-bounded cases.
The blister \hiireg\ region model is a mixture of radiation-bounded and density-bounded nebula, where the nebula is radiation-bounded in the high density zone and the intermediate density cloud, and the nebula is density-bounded in the low density gaseous component.

\section{Conclusion}\label{sec:summary}

\newcode\ is a photoionization code designed for modeling the \hiireg\ regions with arbitrary three-dimensional geometries.
\newcode\ incorporates the Monte Carlo radiative transfer technique with the complete ISM microphysics implemented in \mappings\ code, producing realistic three-dimensional ionization and thermal structures within nebulae.
The accurate cooling functions in \newcode\ promise reliable predictions of the emission-line fluxes.

We put \newcode\ successfully through the Lexington/Meudon benchmarks test, which is a series of artificial photoionized models accounting for the physical conditions of various types of \hiireg\ regions and planetary nebulae.
The emission-line fluxes predicted by \newcode\ are consistent with the fluxes produced by the reference photoionization codes.
We run each Lexington/Meudon benchmark with a high-resolution mode and a low-resolution mode, finding the spatial resolution effect is pronounced for emission-line fluxes produced at the edge of \hiireg\ regions.

We create three fiducial \hiireg\ region models with complex geometries, which are the blister geometry, the bipolar geometry and the fractal geometry.
We find that:
\begin{itemize}
  \item In the blister \hiireg\ region model, the high-density clump is partially ionized.
           The high-density clumpy structure has hotter electron temperature and higher electron density than the low-density gas in the nebula. 
           The diffuse ionized gas is partially ionized and cooler than the average electron temperature of the nebula.            
  \item The bipolar \hiireg\ region model has the similar radial profiles to the spherical model, in terms of the electron temperature, the electron density, the ionizing photon flux, the ionic fraction and the emission-line intensities. 
           However, the bipolar \hiireg\ region has smaller volume size than the spherical model, reducing the integrated emission-line fluxes.
  \item The fractal model has the most complex geometry among the three fiducial \hiireg\ region models.
           Neither the internal ionization and thermal structures nor the integrated emission-line fluxes can be reproduced by simple spherical photoioinzation models.
           The inhomogeneity of ISM density causes the fluctuation of the electron temperature and the electron density.
           The twisted nebular boundary of the fractal model increase the boundary emission-line species.           
\end{itemize}

We demonstrate that \newcode\ is a promising tool for interpreting nebular emission-line behaviors in the era of JWST and the upcoming local integral-field unit surveys, like SDSS-V/LVM (the Local Volume Mapper).

\vspace{25pt}

We are grateful for the valuable comments on this work by an anonymous referee that improved the scientific outcome and quality of the paper. 
This research was conducted on the traditional lands of the Ngunnawal and Ngambri
people.
YFJ is grateful to Brent Groves, Christoph Federrath, Emily Wisnioski and David Nicholls for useful discussions. This research was supported by the Australian Research Council Centre of Excellence for All Sky Astrophysics in 3 Dimensions (ASTRO 3D), through project number CE170100013. L.J.K. gratefully acknowledges the support of an ARC Laureate Fellowship (FL150100113).

\clearpage


\begin{deluxetable}{lccccc}
\tablewidth{8truecm}
\tablecaption {The Lexington/Meudon benchmarks}
\tablehead {
\colhead  { } & HII20 & HII40 & PN150 }
\startdata  
\ $\Phi_{H}~(10^{47} \rm photon~s^{-1})$ & 100.0 & 426.6 &  5.4 \\
\ $T_{eff}~(\rm kK)$ & 20 & 40 & 150   \\
\ $n_{H}~(\rm cm^{-3})$ & 100  & 100 & 3000  \\
\ $R_{in}~(\rm cm)$ & 3$\times\rm10^{18}$ & 3$\times\rm10^{18}$ & 1$\times\rm10^{17}$ \\
\ He & 0.1 & 0.1 & 0.1 \\
\ C   & 2.2$\times\rm10^{-4}$ & 2.2$\times\rm10^{-4}$ & 3.0$\times\rm10^{-4}$ \\
\ N   & 4.0$\times\rm10^{-5}$ & 4.0$\times\rm10^{-5}$ & 1.0$\times\rm10^{-4}$ \\
\ O   & 3.3$\times\rm10^{-4}$ & 3.3$\times\rm10^{-4}$ & 6.0$\times\rm10^{-4}$ \\
\ Ne & 5.0$\times\rm10^{-5}$ & 5.0$\times\rm10^{-5}$ & 1.5$\times\rm10^{-4}$ \\
\ Mg & -- & -- & 3.0$\times\rm10^{-5}$ \\
\ Si   & -- & -- & 3.0$\times\rm10^{-5}$ \\
\ S   & 9.0$\times\rm10^{-6}$ & 9.0$\times\rm10^{-6}$ & 1.5$\times\rm10^{-5}$ \\
\ CPU Time ($33^3$ cells) & 04h14m13s  & 05h28m12s  &  07h03m12s \\
\ CPU Time ($55^3$ cells) & 23h55m33s  & 19h56m32s  &  29h53m25s \\
\enddata  
\end{deluxetable}\label{tab:benchmark}

\begin{table*}
\begin{center}
\caption{Meudon/Lexington H~{\sc ii} Region Benchmark Results: HII20. The \newcode ($33^3$) and \newcode ($55^3$) are the results of the low-resolution model and the high-resolution model respectively. $\Delta\%$ is the deviation of the results between \newcode ($33^3$) and \newcode ($55^3$). ``Med'' column shows the medium value and 1-$\sigma$ deviation of the results given by  {\sc mappings}, {\sc cloudy}, {\sc mocassin3d}, RR and PH codes.}
\begin{minipage}{13.5cm}
\label{tab:HII20}
\begin{tabular}{@{}lccccccccc}
\hline
Line/\hb & \newcode ($33^3$) & \newcode ($55^3$) & $\Delta\%$ & {\sc mappings} & {\sc cloudy} & {\sc mocassin3d} & RR & PH & Med$\pm\sigma$  \\
\hline
\hb\ ($\rm 10^{36}erg~s^{-1}$) & 4.81    & 4.81      &  0.00 & 5.04     & 4.85     & 4.97     & 4.89     & 4.93     & 4.89$\pm$0.09 \\ 
He~{\sc i} 5876                        & 0.0071 & 0.0071 &  0.00 & 0.0110 & 0.0072 & 0.0065 & --          & 0.0074 &  0.0072$\pm$0.0016  \\ 
C~{\sc ii]} 2325+                      & 0.059   & 0.059   &  0.00 & 0.038   & 0.054   & 0.042   & 0.063   & 0.060   &  0.059$\pm$0.010 \\ 
{[N~\sc ii]} 122$\mu$m            & 0.072   & 0.072   &  0.00 & 0.071   & 0.068   & 0.071   & 0.071   & 0.072    & 0.071$\pm$0.001 \\ 
{[N~\sc ii]} 6584+6548             & 0.783   & 0.783   &  0.00 & 0.803   & 0.745   & 0.846   & 0.915   & 0.843     &  0.803$\pm$0.056\\ 
{[N~\sc ii]} 5755                       & 0.0028 & 0.0028 &  0.00 & 0.0030 & 0.0028 & 0.0025 & 0.0033 & 0.0033   & 0.0028$\pm$0.0003  \\ 
{[N~\sc iii]} 57.3$\mu$m          & 0.0028 & 0.0027  &  3.57 & 0.0020 & 0.0040 & 0.0019 & 0.0022 & 0.0031  &  0.0027$\pm$0.0007\\ 
{[O~\sc i]} 6300+6363             & 0.0239  & 0.0241 &  0.84 & 0.0050 & 0.0080 & 0.0088 & --          & 0.0047  & 0.0088$\pm$0.0091 \\ 
{[O~\sc ii]} 7320+7330             & 0.0064  & 0.0064 &  0.00 & 0.0080 & 0.0087 & 0.0064 & 0.0100 & 0.0103  & 0.0080$\pm$0.0017\\ 
{[O~\sc ii]} 3726+3729             & 1.24     & 1.24      &  0.00 & 1.08     & 1.01     & 0.909   & 1.17     & 1.22     & 1.17$\pm$0.13 \\ 
{[O~\sc iii]} 52+88$\mu$m       & 0.0030 &  0.0029 &  3.33  & 0.0020 & 0.0030 & 0.0022 & 0.0017 & 0.0037     & 0.0029$\pm$0.0007\\ 
{[O~\sc iii]} 5007+4959            & 0.0019 & 0.0019  &  0.00  & 0.0010 & 0.0021 & 0.0011 & 0.0010 & 0.0014    & 0.0014$\pm$0.0005 \\ 
{[Ne~\sc ii]} 12.8$\mu$m         & 0.296   & 0.296   &  0.00  & 0.286   & 0.264   & 0.295   & 0.290   & 0.271    &  0.290$\pm$0.013\\ 
{[S~\sc ii]} 6716+6731             & 0.528   & 0.529   & 0.19  & 0.435   & 0.499   & 0.486   & 0.492   & 0.555    &   0.499$\pm$0.039\\ 
{[S~\sc ii]} 4068+4076             & 0.015   & 0.016   & 6.67  & 0.012   & 0.022   & 0.013   & 0.015   & 0.017    &  0.015$\pm$0.003 \\ 
{[S~\sc iii]} 18.7$\mu$m          & 0.324   & 0.324   &  0.00  & 0.398   & 0.445   & 0.371   & 0.374    & 0.365    &  0.371$\pm$0.042 \\ 
{[S~\sc iii]} 9532+9069            & 0.442   & 0.441   &  0.23  & 0.604   & 0.501   & 0.526   & 0.551    & 0.549    &   0.526$\pm$ 0.060\\ 
\hline
\end{tabular}
\end{minipage}
\end{center}
\end{table*}

\begin{table*}
\begin{center}
\caption{Meudon/Lexington H~{\sc ii} Region Benchmark Results: HII40.}
\begin{minipage}{13.5cm}
\label{tab:HII40}
\begin{tabular}{@{}lcccccccccc}
\hline
Line/\hb & \newcode ($33^3$) & \newcode ($55^3$) & $\Delta\%$ & {\sc mappings} & {\sc cloudy} & {\sc mocassin3d} & RR & PH & Med$\pm\sigma$   \\
\hline
\hb\ ($\rm 10^{37}erg~s^{-1}$) & 2.00  & 1.98    & 1.00    & 2.07     & 2.06     & 2.02     & 2.05     & 2.05   & 2.05$\pm$0.03 \\
He~{\sc i} 5876                        & 0.117 & 0.115  &  1.71    & 0.116   & 0.119    & 0.114   & --         & 0.118 & 0.116$\pm$0.047 \\
C~{\sc ii]} 2325+                      & 0.170 & 0.164  &  3.53   & 0.096   & 0.157   & 0.148   & 0.178   & 0.166 & 0.164$\pm$0.027 \\
C~{\sc iii]} 1907+1909             & 0.076 & 0.076  &  0.00   & 0.066   & 0.071   & 0.041   & 0.074   & 0.060  & 0.071$\pm$0.013\\
{[N~\sc ii]} 122$\mu$m            & 0.030 & 0.029  &  3.33  & 0.035   & 0.027   & 0.036   & 0.030   & 0.032   &0.030$\pm$0.003\\
{[N~\sc ii]} 6584+6548             & 0.647 & 0.632  &  2.32   & 0.723   & 0.669   & 0.852   & 0.807   & 0.736  &0.723$\pm$0.082\\
{[N~\sc ii]} 5755                       & 0.0053 & 0.0053 &  0.00     & 0.0050 & 0.0050 & 0.0061 & 0.0068 & 0.0064  & 0.0053$\pm$0.0007\\
{[N~\sc iii]} 57.3$\mu$m          & 0.294   & 0.297   &  1.02   & 0.273   & 0.306   & 0.223   & 0.301   & 0.292       &0.294$\pm$0.029\\
{[O~\sc i]} 6300+6363             & 0.0425 & 0.0290 &  31.76   & 0.0070 & 0.0094 & 0.0065  & --   & 0.0059  &  0.0094$\pm$0.015\\
{[O~\sc ii]} 7320+7330            & 0.020   & 0.027   &  35.00   & 0.024    & 0.029   & 0.025   & 0.036   & 0.032  &0.027$\pm$0.005\\
{[O~\sc ii]} 3726+3729            & 2.08     & 2.05     &  1.44   & 1.88      & 1.94     & 1.92     & 2.26     & 2.19  & 2.05$\pm$0.14\\
{[O~\sc iii]} 52+88$\mu$m      & 2.52     & 2.55     &  1.19   & 2.29      & 2.35     & 2.28     & 2.34     & 2.34  & 2.34$\pm$0.11\\
{[O~\sc iii]} 5007+4959           & 2.35     & 2.37     &  0.85   & 2.17      & 2.21     & 1.64     & 2.08     & 1.93  & 2.17$\pm$0.26\\
{[O~\sc iii]} 4363                     & 0.00419 & 0.00421 &  2.68   & 0.0040 & 0.00235 & 0.0022 & 0.0035 & 0.0032  & 0.0035$\pm$0.0008\\
{[Ne~\sc ii]} 12.8$\mu$m        & 0.202   & 0.199    &  1.49   & 0.217    & 0.177   & 0.212   & 0.196   & 0.181 & 0.199$\pm$0.015 \\
{[Ne~\sc iii]} 15.5$\mu$m       & 0.287   & 0.290    &  1.05   & 0.350    & 0.294   & 0.267   & 0.417   & 0.429  & 0.294$\pm$0.066\\
{[Ne~\sc iii]} 3869+3968         & 0.068   & 0.069    &  1.47   & 0.083    & 0.084   & 0.053   & 0.086   & 0.087  & 0.083$\pm$0.013\\
{[S~\sc ii]} 6716+6731            & 0.251   & 0.194    &  22.71 & 0.133    & 0.137   & 0.141   & 0.130   & 0.155  &  0.141$\pm$0.045 \\
{[S~\sc ii]} 4068+4076            & 0.0108     & 0.0086    &  20.37  & 0.005    & 0.0093 & 0.0060 & 0.0060 & 0.0070 &  0.0070$\pm$0.0021 \\
{[S~\sc iii]} 18.7$\mu$m          & 0.526   & 0.534    &  1.52   & 0.567   & 0.627   & 0.574   & 0.580   & 0.556 &  0.567$\pm$0.033 \\
{[S~\sc iii]} 9532+9069            & 1.02     & 1.04      &  1.96   & 1.25      & 1.13     & 1.21     & 1.28     & 1.23  & 1.21$\pm$0.10\\
\hline
\end{tabular}
\end{minipage}
\end{center}
\end{table*}

\begin{table*}
\begin{center}
\caption{Meudon/Lexington H~{\sc ii} Region Benchmark Results: PN150.}
\begin{minipage}{13.5cm}
\label{tab:PN150}
\begin{tabular}{@{}lccccccccc}
\hline
Line/\hb & \newcode\ ($33^3$) & \newcode\ ($55^3$) & $\Delta\%$ & {\sc mappings} & {\sc cloudy} & {\sc mocassin3d} & PH & Med$\pm\sigma$  \\
\hline
\hb\ ($\rm 10^{35}erg~s^{-1}$) & 2.59  & 2.59    &  0.00  & 2.64    & 2.86      & 2.79     & 2.68    &  2.68$\pm$0.11 \\
He~{\sc i} 5876                        & 0.099 & 0.098  &  1.01  & 0.095   & 0.110    & 0.104   & 0.096 & 0.099$\pm$0.006\\
He~{\sc ii} 4686                       & 0.314 & 0.314  &   0.00   & --         & 0.324    & 0.333   & 0.333  & 0.324$\pm$ 0.010\\
C~{\sc ii]} 2325+                      & 0.215 & 0.216  &   0.47   & 0.141  & 0.277    & 0.141   & 0.450  & 0.216$\pm$0.115\\
C~{\sc ii} 1335                         & 0.015 & 0.014  &  6.67    & --         & 0.119    & 0.121   & 0.119  &0.119$\pm$0.058\\
C~{\sc iii]} 1907+1909             & 2.18   & 2.23    &  2.29    & 1.89    & 1.68      & 1.72     & 1.74    &1.89$\pm$0.24\\
C~{\sc iv} 1549+                      & 2.22   & 2.25    &  1.35    & 3.12    & 2.14      & 2.71     & 2.09   & 2.25$\pm$0.41\\
{[N~\sc i]} 5200+5198              & 0.032 & 0.024  &  25.00    & 0.005  & 0.013    & 0.0067  & 0.020  & 0.020$\pm$0.0105\\
{[N~\sc ii]} 6584+6548             & 1.18   & 1.07    &   9.32    & 1.17    & 1.15      & 1.43      & 1.35  & 1.18$\pm$0.14\\
{[N~\sc ii]} 5755                       & 0.018 &  0.017 &   5.55   & 0.016  & 0.017    & 0.022    & 0.023  & 0.018$\pm$0.003\\
{N~\sc ii]} 1749+                      & 0.102 &  0.099 &   2.94   & 0.091  & 0.106    & 0.111    & 0.139   & 0.106$\pm$0.017\\
{[N~\sc iii]} 57.3$\mu$m          & 0.130 & 0.133  &   2.31   & 0.126  & 0.129    & 0.120    & 0.135  & 0.130$\pm$0.005\\
{N~\sc iv]} 1487+                     & 0.253 & 0.255  &  0.79    & 0.168  & 0.199    & 0.162    & 0.141  & 0.199$\pm$0.048\\
{N~\sc v} 1240+                      & 0.147  & 0.144  &  2.04    & 0.248  & 0.147    & 0.147    & 0.107  & 0.147$\pm$0.047\\
{[O~\sc i]} 63.1$\mu$m           & 0.038  & 0.028  &  26.31    & 0.049  & 0.024    & 0.010    & 0.007  & 0.028$\pm$0.016\\
{[O~\sc i]} 6300+6363             & 0.258  & 0.154  &  40.31    & 0.101  & 0.144    & 0.163    & 0.104  & 0.154$\pm$0.057\\
{[O~\sc ii]} 3726+3729            & 2.38     & 2.28   &  4.20    & 1.75    & 2.03      & 2.24      & 2.66  & 2.28$\pm$0.31\\
{[O~\sc iii]} 51.8$\mu$m          & 1.37    & 1.40   &  2.19    & 1.28    & 1.30      & 1.50      & 1.39  &  1.39$\pm$0.08\\
{[O~\sc iii]} 88.3$\mu$m          & 0.274  & 0.279 &  1.82    & 0.252  & 0.261    & 0.296    & 0.274 & 0.274$\pm$0.015 \\
{[O~\sc iii]} 5007+4959            & 22.0    & 22.4   &  1.82    & 16.8    & 21.4       & 22.63    & 20.8  & 22.00$\pm$2.16\\
{[O~\sc iii]} 4363                      & 0.168  & 0.171 &  1.79    & 0.109   & 0.152     & 0.169   & 0.155  & 0.168$\pm$0.023\\
{[O~\sc iv]} 25.9$\mu$m          & 4.09    & 4.13   &  0.98    & 4.05     & 3.45      & 3.68      & 4.20  & 4.09$\pm$0.30\\
{O~\sc iv]} 1403+                     & 0.181  & 0.182  & 0.55     & --         & 0.183    & 0.203     & 0.225  & 0.183$\pm$0.019 \\
{O~\sc v]} 1218+                      & 0.198  & 0.193  & 2.53     & 0.213   & 0.165    & 0.169     & 0.097  & 0.193$\pm$0.041\\
{[Ne~\sc ii]} 12.8$\mu$m         & 0.055  & 0.042   & 23.64     & 0.043   & 0.028    & 0.030      & 0.027  & 0.042$\pm$0.011\\
{[Ne~\sc iii]} 15.5$\mu$m        & 1.90    & 1.89     & 0.53     & 2.71     & 1.88      & 2.02       & 2.76    & 2.02$\pm$0.42\\
{[Ne~\sc iii]} 3869+3968          & 2.33    & 2.34     & 0.43     & 2.56     & 2.64      & 2.63       & 3.04    & 2.63$\pm$0.26\\
{[Ne~\sc iv]} 2423+                  & 0.663  & 0.668   & 0.75     & 0.832   & 0.707   & 0.749      & 0.723  &  0.723$\pm$0.062\\
{[Ne~\sc v]} 3426+3346          & 0.195   & 0.191   & 2.05     & 0.591   & 0.721    & 0.692     & 0.583  & 0.591$\pm$0.241\\
{[Ne~\sc v]} 24.2$\mu$m        & 0.304   & 0.301   & 0.99     & 0.195   & 0.997    & 1.007     & 0.936  & 0.936$\pm$0.393\\
{Mg~\sc ii} 2798+                    & 2.34     & 2.24     & 4.27     &  0.863  & 2.22      & 2.32       & 0.555  & 2.240$\pm$0.818\\
{[Mg~\sc iv]} 4.49$\mu$m       & 0.136   &  0.137  & 0.74     & 0.115    & 0.121    & 0.111     & 0.042  & 0.121$\pm$0.035\\
{Si~\sc ii]} 2335+                     & 0.152   & 0.170   & 11.84     & 0.127    & 0.160    & 0.160    & --  & 0.160$\pm$0.016\\
{Si~\sc iii]} 1892+                    & 0.179   &  0.185  & 3.35     & 0.083    & 0.446    & 0.325    & 0.382  & 0.325$\pm$0.139\\
{Si~\sc iv]} 1397+                    & 0.141   &  0.143  & 1.42     & 0.122    & 0.183    & 0.214    & 0.172  & 0.172$\pm$0.034\\
{[S~\sc ii]} 6716+6731             & 0.382   &  0.311  & 18.59     & 0.322   & 0.359     & 0.357    & 0.451  & 0.359$\pm$0.050\\
{[S~\sc ii]} 4069+4076             & 0.057   &  0.049  & 14.04     & 0.050   & 0.073     & 0.064    & 0.077  & 0.064$\pm$0.012\\
{[S~\sc iii]} 18.7$\mu$m          & 0.527   &  0.526  & 0.19     & 0.578   & 0.713     & 0.495    & 0.488  & 0.527$\pm$0.084\\
{[S~\sc iii]} 33.6$\mu$m          & 0.202   &  0.200  & 0.99     & 0.240   & 0.281     & 0.210    & 0.206  & 0.210$\pm$0.032\\
{[S~\sc iii]} 9532+9069            & 2.01     &  2.02    & 0.50     & 2.04     & 2.07       & 1.89      & 1.90  & 2.02$\pm$0.08\\
{[S~\sc iv]} 10.5$\mu$m          & 2.69     &  2.72    & 1.12     & 2.25     & 2.09       & 2.25      & 2.22   & 2.25$\pm$0.27\\

\hline
\end{tabular}
\end{minipage}
\end{center}
\end{table*}


\clearpage




\end{CJK*}
\end{document}